\documentclass[
  journal=pasa,
  manuscript=research-paper, 
  year=2020,
  volume=37,
]{cup-journal}

\usepackage{lineno}
\usepackage[breaklinks=true]{hyperref}
\hypersetup{colorlinks,linkcolor=blue,citecolor=blue,urlcolor=blue}
\usepackage{natbib}
\bibpunct{(}{)}{;}{a}{}{,}
\usepackage{graphicx}
\usepackage{txfonts}
\usepackage{lscape}
\usepackage{longtable}
\usepackage{multirow}
\usepackage{multicol}
\usepackage{amssymb}
\usepackage{pifont}
\usepackage{geometry}
\usepackage{pdflscape}
\usepackage{diagbox}
\DeclareUnicodeCharacter{2212}{-}
\usepackage{color,soul}

\usepackage{microtype,siunitx,booktabs}
\title{Infrared study of the star-forming region associated with the UC\,HII regions G45.07+0.13 and G45.12+0.13}

\author{N. Azatyan}
\affiliation{Byurakan Astrophysical Observatory, 0213, Aragatsotn prov., Armenia}
\email[N. Azatyan]{nayazatyan@bao.sci.am}

\author{E. Nikoghosyan}
\affiliation{Byurakan Astrophysical Observatory, 0213, Aragatsotn prov., Armenia}

\author{H. Harutyunian}
\affiliation{Byurakan Astrophysical Observatory, 0213, Aragatsotn prov., Armenia}

\author{D. Baghdasaryan}
\affiliation{Byurakan Astrophysical Observatory, 0213, Aragatsotn prov., Armenia}

\author{D. Andreasyan}
\affiliation{Byurakan Astrophysical Observatory, 0213, Aragatsotn prov., Armenia}

\doi{10.1017/pasa.2020.32}

\received {dd Mmm YYYY}
\revised  {dd Mmm YYYY}
\accepted {dd Mmm YYYY}
\published{22 September 2020}


\begin{document}

\begin{abstract}
Ultra-compact H\,{\sc ii} (UC\,HII) regions are an important phase in the formation and early evolution of massive stars and a key component of the interstellar medium.
The main objectives of this work are to study the young stellar population associated with the G45.07+0.13 and G45.12+0.13 UC\,HII regions, as well as the interstellar medium in which they are embedded. We determined the distribution of the hydrogen column density (N(H$_2$)) and dust temperature (T$_d$) in the molecular cloud using Modified blackbody fitting on {\itshape Herschel} images obtained in four bands: 160, 250, 350, and 500\,$\mu$m. We used near-, mid-, and far-infrared photometric data to identify and classify the young stellar objects (YSOs). Their main parameters were determined by the radiation transfer models. We also constructed a colour-magnitude diagram and K luminosity functions (KLFs) to compare the parameters of stellar objects with the results of the radiative transfer models. We found that N(H$_2$) varies from $\sim$\,3.0\,$\times$\,10$^{23}$ to 5.5\,$\times$\,10$^{23}$\,cm$^{-2}$ within the G45.07+0.13 and G45.12+0.13 regions, respectively. The maximum T$_d$ value is 35\,K in G45.12+0.13 and 42\,K in G45.07+0.13. T$_d$ then drops significantly from the centre to the periphery, reaching about 18--20\,K at distances of $\sim$\,2.6\,pc and $\sim$\,3.7\,pc from IRAS\,19110+1045 (G45.07+0.13) and IRAS\,19111+1048 (G45.12+0.13), respectively. The gas plus dust mass value included in G45.12+0.13 is $\sim$\,3.4\,$\times$\,10$^5$\,M$_\odot$ and $\sim$\,1.7\,$\times$\,10$^5$\,M$_\odot$ in G45.07+0.13. The UC\,HII regions are connected through a cold (T$_d$\,=\,19\,K) bridge. The radial surface density distribution of the identified 518 YSOs exhibits dense clusters in the vicinity of both IRAS sources. The parameters of YSOs in the IRAS clusters (124 objects) and 394 non-cluster objects surrounding them show some differences. About 75\% of the YSOs belonging to the IRAS clusters have an evolutionary age greater than 10$^6$ years. Their slope $\alpha$ of the KLF agrees well with a Salpeter-type Initial Mass Function (IMF) ($\gamma$=1.35) for a high mass range (O–F stars, $\beta\sim$\,2) at 1\,Myr. The non-cluster objects are uniformly distributed in the molecular cloud, 80\% of which are located to the right of the 0.1 Myr isochrone. The slope $\alpha$ of the KLF of non-cluster objects is 0.55\,$\pm$\,0.09, corresponding better to a Salpeter-type IMF for low-mass objects (G–M stars, $\beta\sim$\,1).
Our results show that two dense stellar clusters are embedded in these two physically connected UC\,HII regions. The clusters include several high- and intermediate-mass zero-age main sequence stellar objects. Based on the small age spread of the stellar objects, we suggest that the clusters originate from a single triggering shock. The extended emission observed in both UC\,HII regions is likely due to the stellar clusters.
\end{abstract}

\section{INTRODUCTION }
\label{sec:int}

Massive stars are generally recognised to form inside dense (n\,>\,10$^3$\,cm$^{-3}$) and cold (T\,$\sim$10--30\,K) compact clumps in giant molecular clouds (GMCs) \citep[e.g.][]{lada03}. Based on results obtained for different massive star-forming regions, star formation in these clumps appears to be triggered by compression from external shocks. Other clusters can be formed by self-gravitation of dense regions in molecular clouds \citep[e.g.][]{elmegreen77, zinnecker93}. If star formation is triggered by compression, the age spread of the new generation of stars should be small, while the age spread of young stellar clusters is expected to be larger in self-initiated condensations \citep{zinnecker93}.

Although few in number relative to solar-mass stars, massive stars and the star clusters hosting them are thought to play an important role in the evolution of galaxies. They affect their environment by shaping the morphology, energy, and chemistry of the interstellar medium (ISM) through phenomena such as outflows, stellar winds, and supernovae \citep{mckee03}. The energy injected into the surrounding neutral ISM may trigger the formation of new stars \citep{elmegreen77} or indeed the opposite effect, inhibiting star formation due to dispersal of molecular clouds \citep{williams97}. 

The study of massive stars involves certain difficulties \citep{zinnecker07,mckee07}. The main challenges in studying high-mass star formation include: (i) the newly formed massive stars are deeply embedded in GMCs, (ii) massive stars are rare, and (iii) they begin burning their nuclear fuel, i.e. reach the zero-age main sequence (ZAMS) while still accreting \citep{mckee03,peters10}. Understanding the formation and early evolution of massive stars requires detailed knowledge of the environments where star-forming events occur, e.g. the main properties of the surrounding material and the characteristics of the embedded stellar population.

Massive stars produce powerful Lyman continuum emission that is sufficiently energetic to ionise their surroundings and create observable ionised H\,{\sc ii} regions. These regions progress from hyper-compact through ultra-compact (UC\,HII) to compact and finally diffuse H\,{\sc ii} region phases, during which they either dissociate or blow out their surrounding medium \citep{churchwell02,keto07}. Thus, UC\,HII regions are thought to be the sites of the early stages of massive star formation. These represent ideal natural laboratories to investigate the influence of hot massive stars on their environment, including the morphology of their surrounding material and the triggering of star formation. 

The UC\,HII regions are dense, compact bubbles of photoionised gas in diameter less than 0.1\,pc \citep{wood89,kurtz00,garay99,churchwell02}, with an estimated age of 10$^4$--10$^5$\,years based on their spatial diameters and typical expansion rates. Most UC\,HII regions have been identified by their centimetre-wavelength free-free emission, as these photons can penetrate the surrounding dust cores and dense molecular gas. A study by \citet{wood89} showed that UC\,HII regions are well distinguished at far- and mid-infrared wavelengths. Mid-infrared observations have also confirmed the presence of low-density extended emissions surrounding UC\,HII regions \citep[e.g.][]{Fuente2020a, Fuente2020b}. The energy flux from the halo is quite large compared to that of small UC\,HII cores; most likely, this requires ionisation by a cluster of hot stars as the energy is usually higher than expected from single high-mass stars \citep{churchwell02}.

To constrain the formation mechanisms of embedded clusters, a detailed study of the stellar and gas content in UC\,HII regions and their immediate neighbourhoods provides vital information. The star-forming regions associated with IRAS\,19110+1045 and IRAS\,19111+1048 sources, referred to as G45.07+0.13 and G45.12+0.13 UC\,HII regions, respectively \citep{wood89}, are part of the Galactic Ring Survey Molecular Cloud (GRSMC) 45.46+0.05 large star formation complex \citep{simon01}, where several other UC\,HII regions are found. This complex is thus an ideal laboratory to investigate the early stages of massive star formations and their influence on natal environments.

From the images obtained at 2 and 6\,cm radio wavelengths, \citet{wood89} determined a spherical morphology of the ionised gas in the G45.07+0.13 region and a cometary shape in G45.12+0.13. In 1.3\,mm continual images, G45.12+0.13 has an elliptical shape, elongated from north-east to south-west \citep{hernandez14}. At MIR wavelengths, both objects show extended emissions and each includes at least 2 UC\,HII regions \citep{Fuente2020a,Fuente2020b}. Given their proximity in the plane of the sky and similar LSR velocities, \citet{hunter97} suggested that G45.12+0.13 and G45.07+0.13 are at the same distance (8.3\,kpc). \citet{simon01} determined a kinematic distance of 6\,kpc for the GRSMC 45.46+0.05 complex containing the two UC\,HII regions. \citet{fish03} and \citet{han15} obtained near and far kinematic distances for both G45.07+0.13 and G45.12+0.13 of $\sim$4.0 and $\sim$8.0\,kpc. The estimated far distance is consistent with the previous estimate by \citet{hunter97} and a more recent one based on the trigonometric parallax method \citep[7.75$\pm$0.45\,kpc,][]{wu19}. Therefore, we adopt a distance of 7.8\,kpc in our study.

Multi-wavelength studies suggest that both regions are sites of active massive star formation, where IRAS\,19111+ 1048 (G45.12+0.13) is in a more advanced stage \citep{hunter97,vig06}. \citet{hunter97} were the first to detect outflows from these sources. Their CO\,($J$\,=\,$6-5$) map shows bipolar outflows with an origin well centred on the radio position of both UC\,HII regions. The continuum-subtracted molecular hydrogen image in \citet{persi19} shows a pair of faint H$_2$ emission knots in G45.07+0.13. Both regions contain type-I OH masers \citep{argon00}, however, only G45.07+0.13 produces H$_{2}$O \citep{hofner96} and methanol maser emissions \citep{hernandez19,breen19}. The probe of massive star-forming clumps, SO $J_{k}$\,=\,$1_{0}-0_{1}$ (i.e. 30\,GHz) emission, has also been detected in both regions \citep{mateen06}.

The NIR data, as well as high-resolution radio measurements, enabled \citet{vig06} and \citet{Rivera2010} to conclude that the IRAS\,19111+1048 region contains numerous ZAMS stars energising a compact and evolved H\,{\sc ii} region. \citet{liu19} reached the same conclusion based on multi-wavelength images observed with SOFIA-FORCAST at wavelengths from $\sim$10\,$\mu$m to 40\,$\mu$m. Based on integrated radio emissions, these sources' spectral types are O6\,-\,BO and the power-law index of the IMF is 5.3$\pm$0.5 for the mass range 14<m/M$_{sun}$<33 \citep{vig06}. \citet{azatyan16} revealed a compact group of pre-main-sequence (PMS) objects ($\sim$50 members) in the vicinity of IRAS\,19110+1045.

In this paper, we present the results of a near-, mid-, and far-infrared (NIR, MIR, and FIR) study of UC\,HII regions G45.07+0.13 and G45.12+0.13. We aim to better understand (i) the physical properties of dense molecular and ionised gas in the immediate neighbourhood of UC\,HII regions; (ii) the properties of embedded massive stars or star clusters; (iii) whether the formation of the embedded massive stars or star clusters was triggered; and (iv) whether the presence of a new generation of stars was triggered by the activity of the embedded star(s).

We have organised the paper as follows. Section \ref{2} describes the used data; in Section \ref{3.1} we discuss the properties of the gas and dust and from Section \ref{3.2} to Section \ref{3.6}, we analyse the stellar population in the region. Finally, the study results are summarised in Section \ref{5}.

\section{Used Data}
\label{sec:obs}

\label{2}
In our study, we used data covering a wide range of NIR to FIR wavelengths. The first dataset comprises archival NIR photometric data in the J, H, and K bands of the Galactic Plane Survey DR6 \citep[UKIDSS GPS,][]{lucas08} with a resolution of 0.1"/px. This survey is complete to approximately 18\,mag in the K band and provides a percentage probability of an individual object being a star, galaxy, or noise.

Archival MIR observations were obtained from the Galactic Legacy Infrared Midplane Survey Extraordinaire \citep[GLIMPSE,][]{churchwell09}, using the {\itshape Spitzer} Infrared Array Camera \citep[IRAC,][]{fazio04}. The four IRAC bands are centred at approximately 3.6, 4.5, 5.8, and 8.0\,$\mu$m with a resolution of 0.6"/px. At longer wavelengths, we used data from a survey of the inner Galactic plane using the Multiband Infrared Photometer for {\itshape Spitzer} (MIPSGAL). The survey field was imaged in 24 and 70\,$\mu$m passbands with resolutions of 6"/px and 18"/px, respectively \citep{carey09}; however, only 24\,$\mu$m data were available for the studied star-forming region. The point source photometric data in this star-forming region were downloaded from the NASA/IPAC Infrared Science Archive. We also used Wide-field Infrared Survey Explorer \citep[WISE,][]{wright10} data in the 3.4, 4.6, 12, and 22\,$\mu$m bandpasses and Midcourse Space Experiment \citep[MSX,][]{price01} full-plane survey data in the 8.28, 12.13, 14.65, and 21.3\,$\mu$m bands, which are accessible through VizieR.

To study gas and dust, as well as deeply embedded point sources, we used FIR observations, in the 70–500\,$\mu$m range, obtained with the Photodetector Array Camera and Spectrometer \citep[PACS,][]{poglitsch10} and the Spectral and Photometric Imaging Receiver \citep[SPIRE,][]{griffin10} on board the 3.5\,m {\itshape Herschel} Space Observatory \citep{pilbratt10}. For our analyses, we used photometric data and images from the PACS 70 and 160\,$\mu$m catalogues, in addition to {\itshape Herschel} infrared Galactic Plane Survey \citep[Hi-GAL,][]{molinari16} data at 70, 160, 250, 350, and 500\,$\mu$m. The corresponding {\itshape Herschel} half-power beamwidth (HPBW) values are 5.0"\, at 70\,$\mu$m, 11.4"\, at 160\,$\mu$m, 17.8"\, at 250\,$\mu$m, 25.0"\, at 350\,$\mu$m, and 35.7"\, at 500\,$\mu$m. The point and extended source photometry were downloaded from the NASA/IPAC Infrared Science Archive. For objects with measured fluxes in both the PACS and Hi-GAL catalogues, we selected the Hi-GAL photometric data.

\begin{figure*}[hbt!]
\centering
\includegraphics[width=0.9\linewidth]{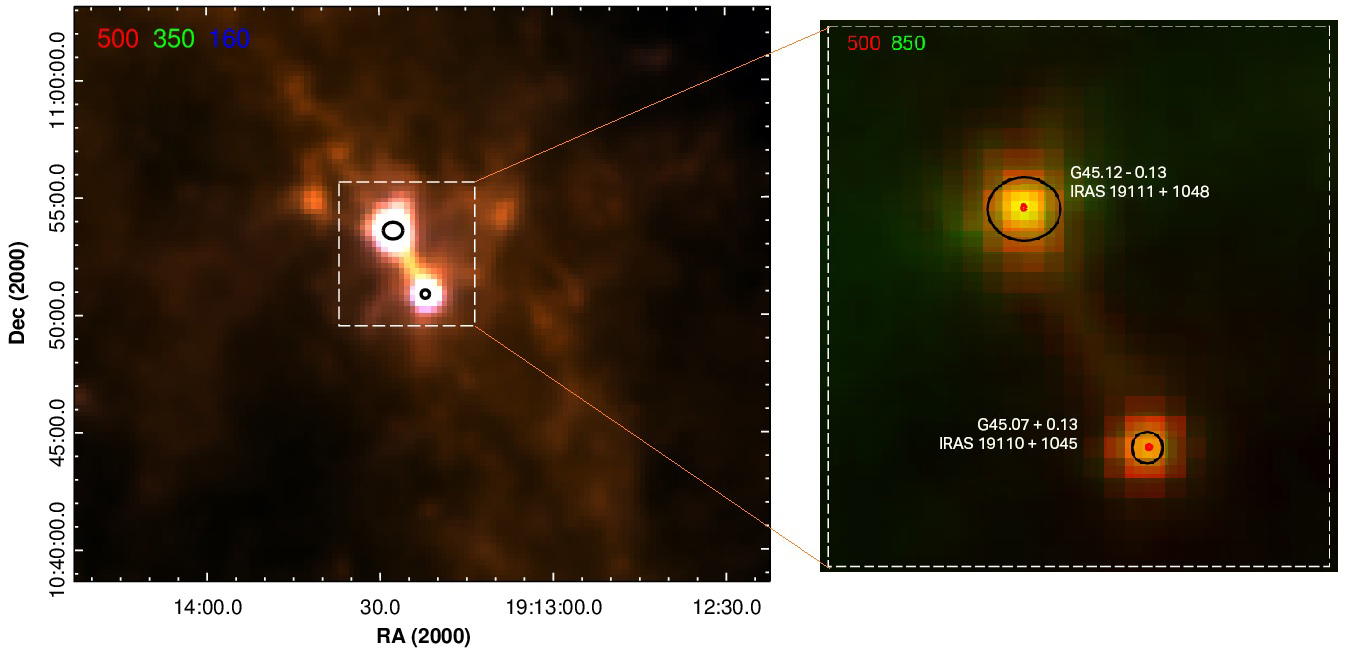}
\caption{Colour-composite images of the  G45.12+0.13 and G45.07+0.13 UC\,HII regions. {\itshape Left panel}: {\itshape Herschel} 160\,$\mu$m (blue), 350\,$\mu$m (green), and 500\,$\mu$m (red); {\itshape right panel}: the zoomed area (white dotted square) at SCUBA 850\,$\mu$m (green) and {\itshape Herschel} 500\,$\mu$m (red). The positions and dimensions of the radio sources obtained at 2- and 6-cm wavelengths in \citet{wood89} are marked by black circles. A red dot represents the position of an IRAS source.}
\label{fig:1}
\end{figure*}

\begin{figure*}[hbt!]
\centering
\includegraphics[width=10cm]{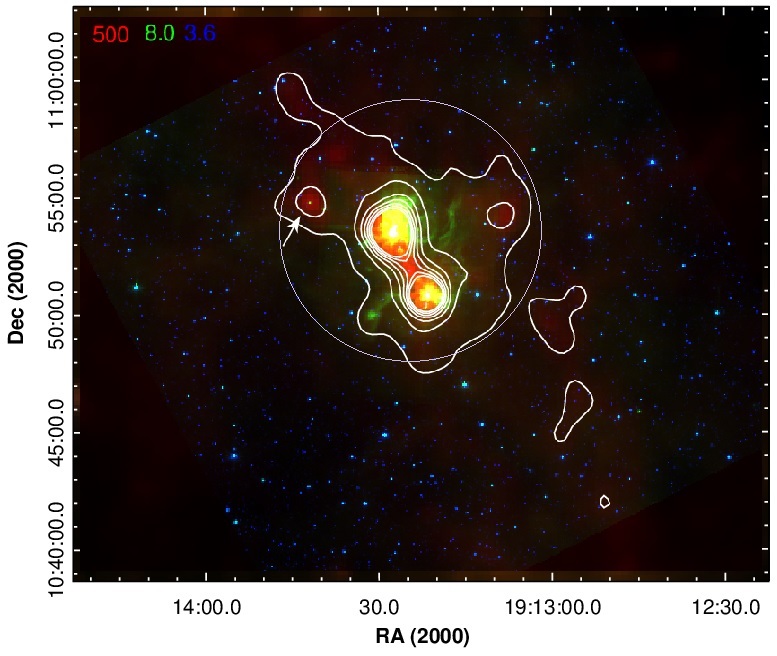}
\caption{Colour-composite image of the G45.12+0.13 and G45.07+0.13 UC\,HII regions: {\itshape Spitzer} IRAC 3.6\,$\mu$m (blue), 8.0\,$\mu$m (green) and {\itshape Herschel} 500\,$\mu$m (red). The latter is also shown with a white isophotes. The external isophot corresponds to a surface brightness of 0.55\,Jy/pix, which exceeds the background $\sigma$ = 0.18\,Jy/pix by threefold. Difference between the isophotes level is 0.25\,Jy/pix. The white arrow shows a small clumpy structure around the MSX G045.1663+00.0910 object. The circle outlines an area with a radius of 6\,arcmin, which almost completely covers the region of study.} 
\label{fig:2}
\end{figure*}

\section{Results and Discussion}
\label{3}

\subsection{Dust emission}
\label{3.1}
The Hi-GAL observations provide a complete and unbiased view of continuum emission in the Galactic plane in 70, 160, 250, 350, and 500\,$\mu$m bands. This wavelength range covers the peak of the spectral energy distribution (SED) of cold dust emission, constraining important ISM parameters such as the hydrogen column density (N(H$_2$)) and the dust temperature (T$_d$).

In Figure \ref{fig:1}, we present a colour-composite image (left panel) of three Hi-GAL bands (PACS 160\,$\mu$m and SPIRE 350 and 500\,$\mu$m) covering the molecular cloud. The G45.12+ 0.13 and G45.07+0.13 UC\,HII regions stand out sharply in terms of brightness. The black circles coincide with the positions and dimensions of radio continuum radiation \citep{wood89}. These sources align well with the brightest part of the 850\,$\mu$m emission (right panel). Figure \ref{fig:1} shows that both UC\,HII regions are embedded in a molecular cloud, with clear clump-like and filamentary structures. The images also indicate that the UC\,HII regions are connected by a relatively colder bridge and are thus very likely a physically bound system. 

Figure \ref{fig:2} shows a colour-composite image of {\itshape Spitzer} IRAC 3.6 and 8.0\,$\mu$m, and SPIRE 500\,$\mu$m bands, illustrating that the UC\,HII regions are also very bright in the MIR wavelength range. This indicates a relatively high ISM temperature in the vicinity of the UC\,HIIs and, therefore, the presence of a significant number of hot stars \citep{vig06,Rivera2010}. This is also confirmed by the stellar population study results in this work (see the text below). Furthermore, the MIR image is saturated in the centre of both regions, which is the case for several UC\,HIIs \citep{churchwell02, Fuente2020a}. In Figure \ref{fig:2}, the arrow shows a small clumpy structure coinciding with the MSX G045.1663+00.0910 object. Three objects from the Bolocam Galactic Plane Survey (BGPS) are located within the external {\itshape Herschel}\,500\,$\mu$m contour; of these objects 6535 and 6536 in the Bolocat V2.1 catalogue \citep{ellsworth15} are associated with IRAS\,19110+1045 and IRAS\,19111+1048, respectively. The third object (6737) is associated with MSX G045.1663+00.0910. The distance estimates for these three BGPS objects are almost identical \citep[7.5\,--\,7.9\,kpc,][]{ellsworth15}, consistent with the findings of \citet{wu19} and confirming that they belong to the same star formation region. We also would like to highlight the filamentary structure directed north-east towards the most extended part of the GRSMC 45.46+0.05 molecular cloud, to which UC\,HIIs G45.12+0.13 and G45.07+0.13 belong \citep{simon01}. 

Dust emission in the FIR range can be modelled as a modified blackbody $I_{\nu}=k_{\nu}\mu_{H_2}m_{H}N(H_{2})B_{\nu}(T_{d})$, where $k_{\nu}$ is the dust opacity, $\mu_{H_2}=2.8$ representing the mean weight per hydrogen molecule \citep{kauffmann2008}, $m_{H}$ is the mass of hydrogen, $N(H_{2})$ is the hydrogen column density, and  $B_{\nu}(T_{d})$ is the Planck function at dust temperature $T_{d}$.

The dust opacity in the FIR is usually parameterised as a power-law normalised to the value $k_{0}$ at a reference frequency $\nu_{0}$ such that $k_{\nu}=k_{0}(\nu/\nu_{0})^{\beta}\,cm^{-2}g^{-1}$. Following \citet{hildebrand83,konyves2015}, we adopted $k_{0}=0.1\,cm^{-2}g^{-1}$ at 300\,$\mu$m (i.e. gas-to-dust ratio of 100) and $\beta=2$. \citet{roy2014} showed that this assumption is valid within better than 50\% of the bulk of molecular clouds.
We used four intensity maps from 160\,$\mu$m to 500\,$\mu$m to derive N(H$_{2}$) and T$_{d}$ maps. Based on the discussion in previous studies \citep[e.g.][]{battersby11}, we excluded the 70\,$\mu$m data as the optically thin assumption may not hold. In addition, the emission here would have a significant contribution from a warm dust component, thus, modelling with a single-temperature blackbody would over-estimate the derived temperature. The three shortward intensity maps were degraded to the spatial resolution of the 500\,$\mu$m band; the four maps were then translated to a common coordinate system using 14" pixels at all wavelengths.

\begin{figure*}[hbt!]
\centering
\includegraphics[width=8.9cm]{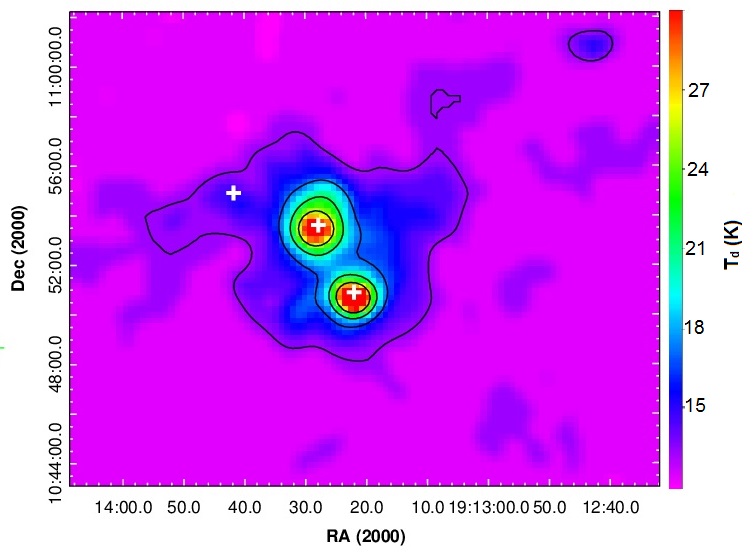}
\includegraphics[width=9.0cm]{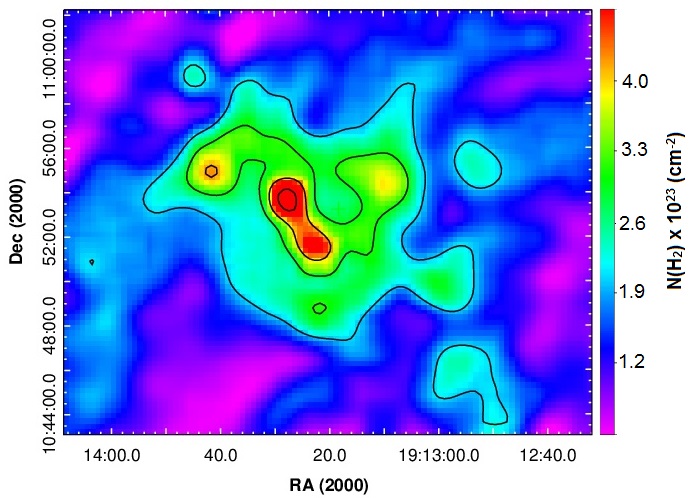}
\caption{Maps of the N(H$_2$) (\textit{left panel}) and T$_d$ (\textit{right panel}) of the region. On the T$_d$ map the outer isotherm corresponds to 13 K and the interval between
isotherms is 4 K. On the N(H$_2$) map the outer isodense corresponds to 2.0\,$\times$\,10$^{23}\,cm^{-2}$ and interval between isodenses is 1.0\,$\times$\,10$^{23}\,cm^{-2}$. The positions of the IRAS and BGPC\,6737 sources are marked by white crosses.}
\label{fig:3}
\end{figure*}

The SED fitting procedure was executed pixel by pixel. Following \citet{pezzuto2021}, we created a grid of models by varying only the temperature in the range $5\leq T_{d}(K)\leq 50$ in steps of 0.01\,K; for each temperature $T_{j}$, the code computes the intensity at FIR wavelengths for all bands. Since $I_{\nu}$ is linear with N(H$_{2}$), we can compute the column density for each pixel using the least-squares technique. The
uncertainty values of $I_{\nu}$ for SPIRE and PACS are 10\%  and 20\%, respectively \citep{konyves2015}.

From the derived column density values, we estimate the mass of the dust clumps using the following expression:
\begin{equation}
    \label{equ:4}
    M_{\rm clump}=\mu_{H_2}m_HArea_{\rm pix}\sum N(H_2) \, ,
\end{equation}
where Area$_{\rm pix}$ is the area of a pixel in cm$^2$, i.e. we calculated the mass in each pixel and then summed over all pixels within the radius. 

The N(H$_2$) and T$_d$ maps of the wider region surrounding the G45.12+0.13 and G45.07+0.13 UC\,HII objects are shown in Figure \ref{fig:3}. Both UC\,HII regions are distinct from the molecular cloud due to their high dust temperature and column density with an almost spherically symmetric distribution. In general, within both regions, T$_{d}$ varies from about 17\,K to 40\,K and N(H$_{2}$)  varies from about $3.0\,\times\,10^{23}\,cm^{-2}$ to 5.5\,$\times$\,10$^{23}$\,cm$^{-2}$, consistent with other UC\,HIIs \citep{churchwell10}.

\textbf{G45.07+0.13.} In this region, the IRAS source is somewhat offset from the density maximum, which is $\sim$\,5.0\,$\times\,10^{23}$~cm$^{-2}$. Near IRAS\,19110+1045, the column density is $\sim$\,3.5\,$\times\,10^{23}$~cm$^{-2}$. This value agrees well with the data obtained in \citet{churchwell10} (i.e. $3.0\,\times\,10^{23}~cm^{-2}$) and is almost an order of magnitude smaller than the value of \citet{hernandez14} ($2.7\,\times\,10^{24}~cm^{-2}$). In both studies, the results were obtained based on millimetre observations. The IRAS source is located close to the dust temperature maximum (T$_d$\,=\,42\,K). This value is consistent with the data obtained in the submillimetre range \citep[T\,=\,42.5\,K,][]{Rivera2010}, and, within error, with the data presented in \citet{hernandez14} (T\,=\,$82^{+25}_{-39}\,K$).
As shown in Fig.~\ref{fig:3}, the dust temperature drops significantly from the centre to the periphery, reaching a value of about 18\,--\,20\,K. Beyond around a 5-pixel ($\sim$\,1.2\'\, or $\sim$\,2.6\,pc) distance from the IRAS source, T$_d$ remains constant. This may correspond to the extent of the region of influence of the ionising source(s). The size of this region correlates with the size of the extended emission \citep[1\'\,$\times\,$1\'\,,][]{Fuente2020a}, and this region's mass is $\sim$\,1.7\,$\times\,$10$^5$\,M$_\odot$.

\textbf{G45.12+0.13.} In this region, the position of IRAS\,19111+ 1048 coincides with the maxima of both the column density (5.5\,$\times$\,10$^{23}$\,cm$^{-2}$) and temperature (35\,K). Our N(H$_2$) value exceeds that presented in \citet{churchwell10} by more than an order of magnitude, however, it is comparable with the value obtained through analysis of the multi-transition $^{13}$CO emission \citep{churchwell92}. The T$_d$ value is in good agreement with the data presented in \citet{Rivera2010} (T\,=\,39\,K). The isotherms are slightly elongated towards the northwest (Fig.~\ref{fig:3}), which may relate to the presence of two UC\,HIIs (G45.12+0.13 and G45.13+0.14) that are separated from each other in a northwest direction \citep{Fuente2020a}. 
Similar to G45.07+0.13, the dust temperature in G45.12+0.13 drops significantly further from the centre, reaching a value of around 18\,--\,20\,K in its periphery. The T$_d$ decline levels out at around 7\,pixels ($\sim$\,1.6\,\'\, or $\sim$\,3.7\,pc) from the IRAS source. As above, the size of this region correlates with the size of the extended emission \citep[{1.5\'\,$\times$\,1.5\'\,,}][]{Fuente2020a}. The larger radius of the ionised region indicates that the physical extent of ionised gas around IRAS\,19111+1048 is larger than that of its neighbour IRAS\,19110+1045 \citep{vig06}. The mass of this region is $\sim$\,3.4\,$\times$10$^5$\,M$_\odot$.

In both regions, the extent of the ionised area significantly exceeds the size of the UC\,HII radio cores. Previous studies have shown that many UC\,HII regions are surrounded by extended emission (on an arc-minute scale) produced by low-density ionised halos \citep[][and ref. therein]{churchwell02, Fuente2020a, Fuente2020b}. The energy flux from these halos is quite large relative to that of the small UC\,HII cores, implying ionisation by a cluster of hot stars rather than a single star \citep{churchwell02}.

In Figure \ref{fig:3}, a region with relatively high density (N(H$_2$)\,$\approx$\, 4.3\,$\times$\,10$^{23}$\,cm$^{-2}$) and low temperature (T$_d$\,$\approx$\,19\,K) is positioned between the two UC\,HII regions. This region coincides with the bridge between the UC\,HII regions, which is also distinguishable in the {\itshape Herschel} and {\itshape Spitzer} images (Figs. \ref{fig:1} and \ref{fig:2}), as well as on the H30$\alpha$ maps \citep{churchwell10}, suggesting that these UC\,HII regions are physically connected.

The distribution of the hydrogen column density in the surrounding molecular cloud has a complex, clumpy structure -- in general, outwards from the UC\,HII regions, N(H$_{2}$) varies from about 3.0\, to $0.5\,\times\,10^{23}\,cm^{-2}$ and T$_{d}$ varies from about 16 to 12\,K. Fig.~\ref{fig:3} shows that the  ISM forms a wel--defined concentration around the BGPS\,6737 object; here, the column density value increases to $\sim\,4.3\,\times\,10^{23}\,cm^{-2}$.

\subsection{Stellar population}
\label{3.2}

\subsubsection{Selection of stellar objects}
\label{3.2.1}

To select and study the potential stellar members of the star-forming region, we focused on the part of the molecular cloud where the surface brightness at 500\,$\mu$m exceeds the background by 3\,$\sigma$. This forms a circle with a 6\,arcmin radius around the geometric centre $\alpha$\,=\,19:13:24.2, $\delta$\,=\,+10:53:38, as shown in Figure \ref{fig:2}. The identification of stellar objects was performed with GPS UKIDSS-DR6 as the main catalogue, and then other MIR and FIR catalogues were cross-matched with it. 

The GPS UKIDSS-DR6 catalogue provides the probability of an object being a star, galaxy, or noise based on its image profile. The UKIDSS team recommends that sources classified as noise should be excluded since most of them are not real sources \citep{lucas08}. However, "galaxies" and "probable galaxies" should be included in the full search, since many of these are unresolved pairs of stars or nebulous stars and could be potential members of the star-forming region. Therefore, we selected objects with a <\,30\,\% probability of being noise and a magnitude of K\,<\,18.02\,mag, taking into account the K band limit of the UKIDSS survey. In addition, we removed objects with zero errors of measured magnitudes in the J, H, and K bands. This yielded a total of approximately 28,000 objects. 

\begin{figure}[hbt!]
\centering
\includegraphics[width=1\linewidth]{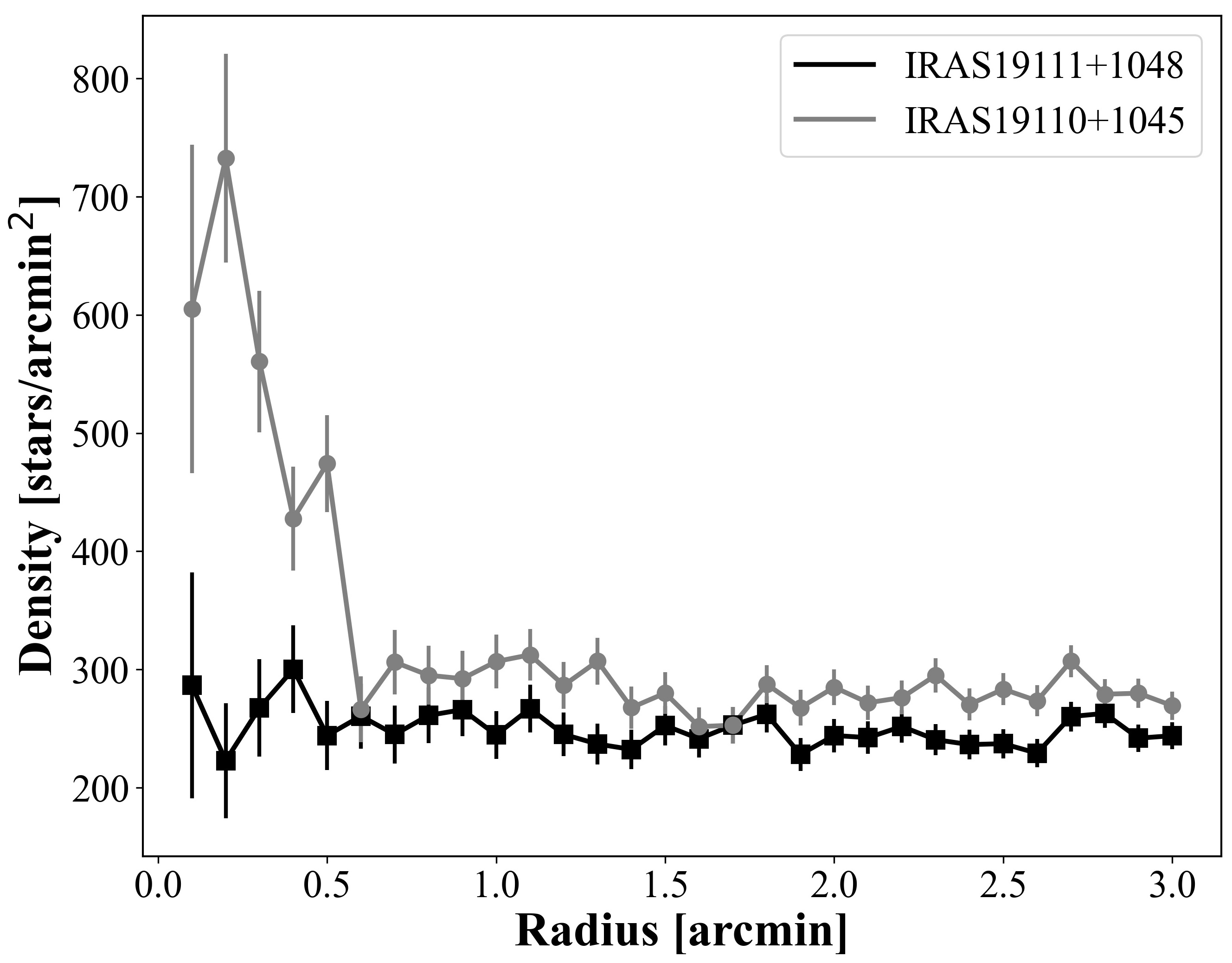}
\caption {Radial distribution of the stellar densities relative to IRAS\,19111+1048 (black line) and IRAS\,19110+1045 (grey line) sources, respectively. Vertical lines are standard errors.}
\label{fig:rVSd}
\end{figure}

To confirm the existence of clusters in the vicinity of both IRAS sources, we constructed the radial density distribution of stars with respect to IRAS\,19110+1045 and IRAS\,19111+1048 (grey line and black line, respectively, Figure \ref{fig:rVSd}). The stellar density was determined in rings of width 0.1\'\, by dividing the number of stars by the surface area. The standard error for the number of stars in each ring is used as a measure of uncertainty. The radial density distribution of the stars shows a well-defined concentration of stars around the IRAS\,19110+1045 source, confirming the existence of a cluster. Starting from a $\sim$\,0.8\,\'\, radius, the stellar density does not exceed the average density of the field. The existence of a cluster around the IRAS\,19111+1048 source is not well-defined and no density concentration was recorded around the MSX G045.1663+00.0910 source.

The MIR and FIR photometric catalogues were cross-matched with the GPS UKIDSS-DR6 catalogue within 3\,$\sigma$ of the combined error--matching radius (Col. 3 in Table \ref{tab: catalogs1}). Matching radii were evaluated considering the positional accuracy of each catalogue (Col. 2 in Table \ref{tab: catalogs1}). Sometimes, the matching radii were quite large, especially for the FIR catalogues. Additionally, multiple NIR objects can be identified by only one FIR object; in such cases, priority was given to the object with significantly higher brightness in the NIR/MIR range. If necessary, we selected the closest one by coordinates. When it was not possible to identify a single object, we did not use those FIR data to avoid possible errors. The set of steps that we used to identify YSOs in the combined photometric catalogue is presented in the following sub-sections.


\begin{table}
\caption{Properties of the catalogues cross-matched with the GPS UKIDSS-DR6}

\resizebox{1\textwidth}{!}{
\label{tab: catalogs1}
\begin{tabular}{l c c l} \toprule
Catalogue name    &       Positional accuracy     &       3\,$\sigma$ of combined error    &   Reference\\
&       (arcsec)       &       (arcsec)       &  \\
\hline\noalign{\smallskip}
(1)     &       (2)     &       (3)     &       (4)     \\
\hline \noalign{\smallskip}
GPS UKIDSS-DR6  &       0.3     &       $-$     &       1 \\
GLIMPSE      &       0.3     &       1.2     &       2    \\
ALLWISE &       1       &       3       &       3    \\
MIPSGAL &       1       &       3       &       4    \\
MSX     &       3.3     &       10      &       5     \\
IRAS    &       16      &       48      &6  \\
PACS: Extended source    &       2.4     &    7.2     &   7 \\
PACS: 70\,$\mu$m&       1.5     &       5       &       8   \\
PACS: 160\,$\mu$m&       1.7     &       5.2     &      9   \\
Hi-GAL: 70, 160, 250,350,500\,$\mu$m&   2   &       6.1     &       10    \\ \bottomrule
\end{tabular}
}
1. \citet{lucas08}; 2. \citet{churchwell09}; 3. \citet{wright10}; 4. \citet{carey09}; 5. \citet{egan03}; 6. \citet{neugebauer84}; 7. \citet{maddox17}; 8. \citet{poglitsch10}; 9. \citet{poglitsch10}; 10. \citet{sanchez14}.\\

\textit{\textbf{Notes.}} (1) Name of used catalogue, (2) Positional accuracy of each catalogue, (3) 3\,$\sigma$ of combined error of the cross-matched catalogues, (4) Source of used data.
\end{table}

\subsubsection{Colour--colour diagrams}
\label{3.2.2}

\begin{figure*}
\centering
\includegraphics[width=0.41\linewidth]{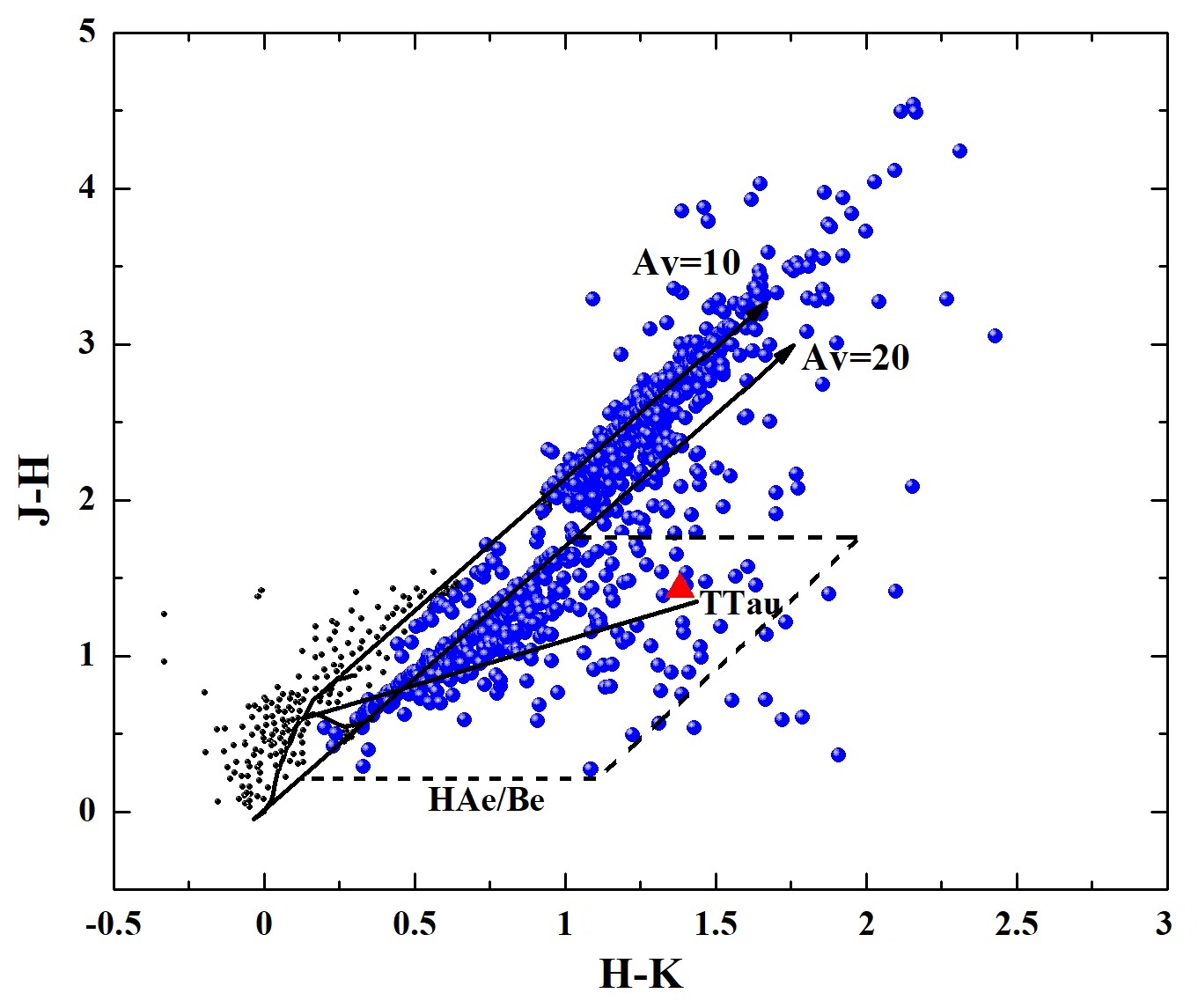}
\includegraphics[width=0.41\linewidth]{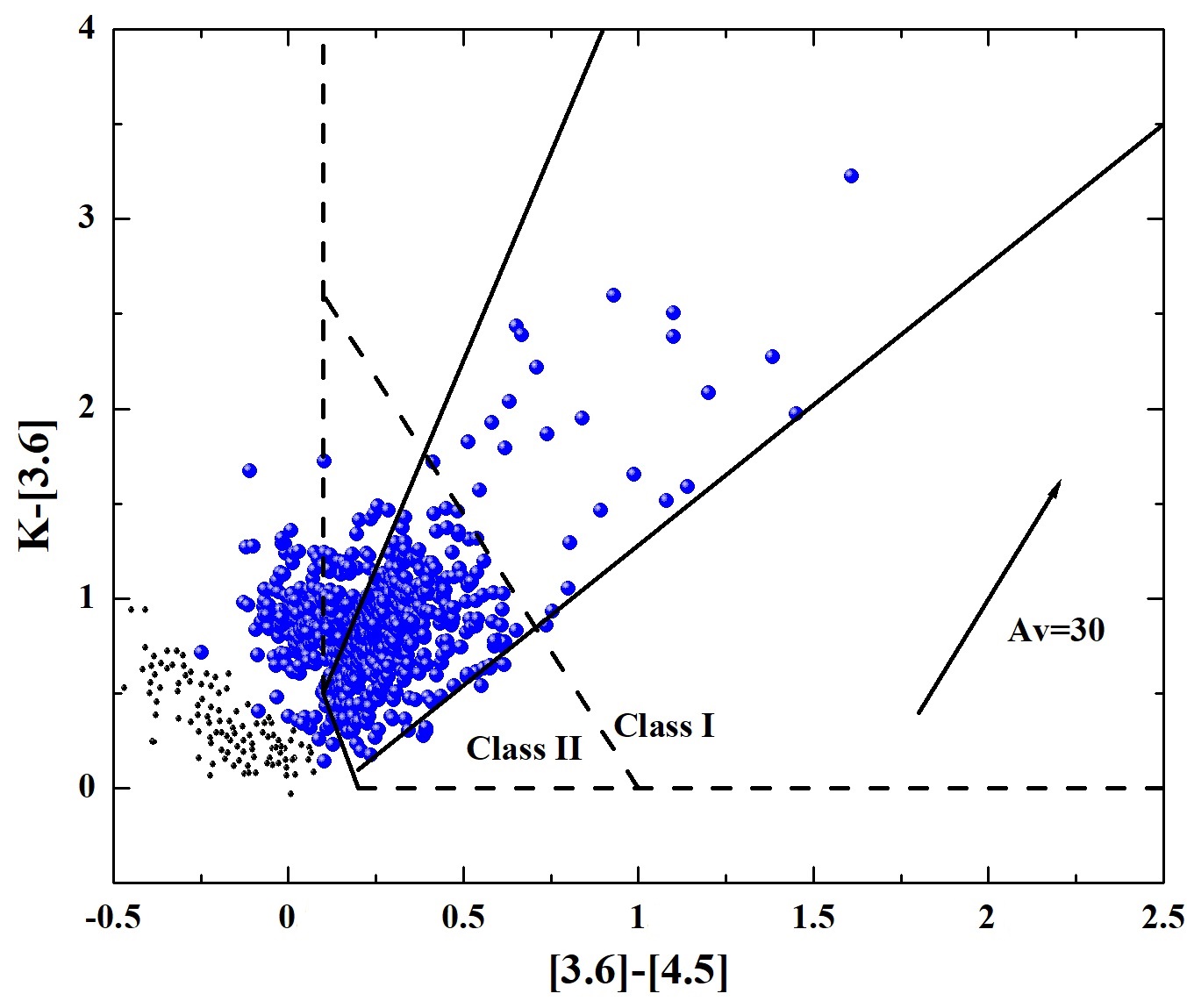}
\includegraphics[width=0.425\linewidth]{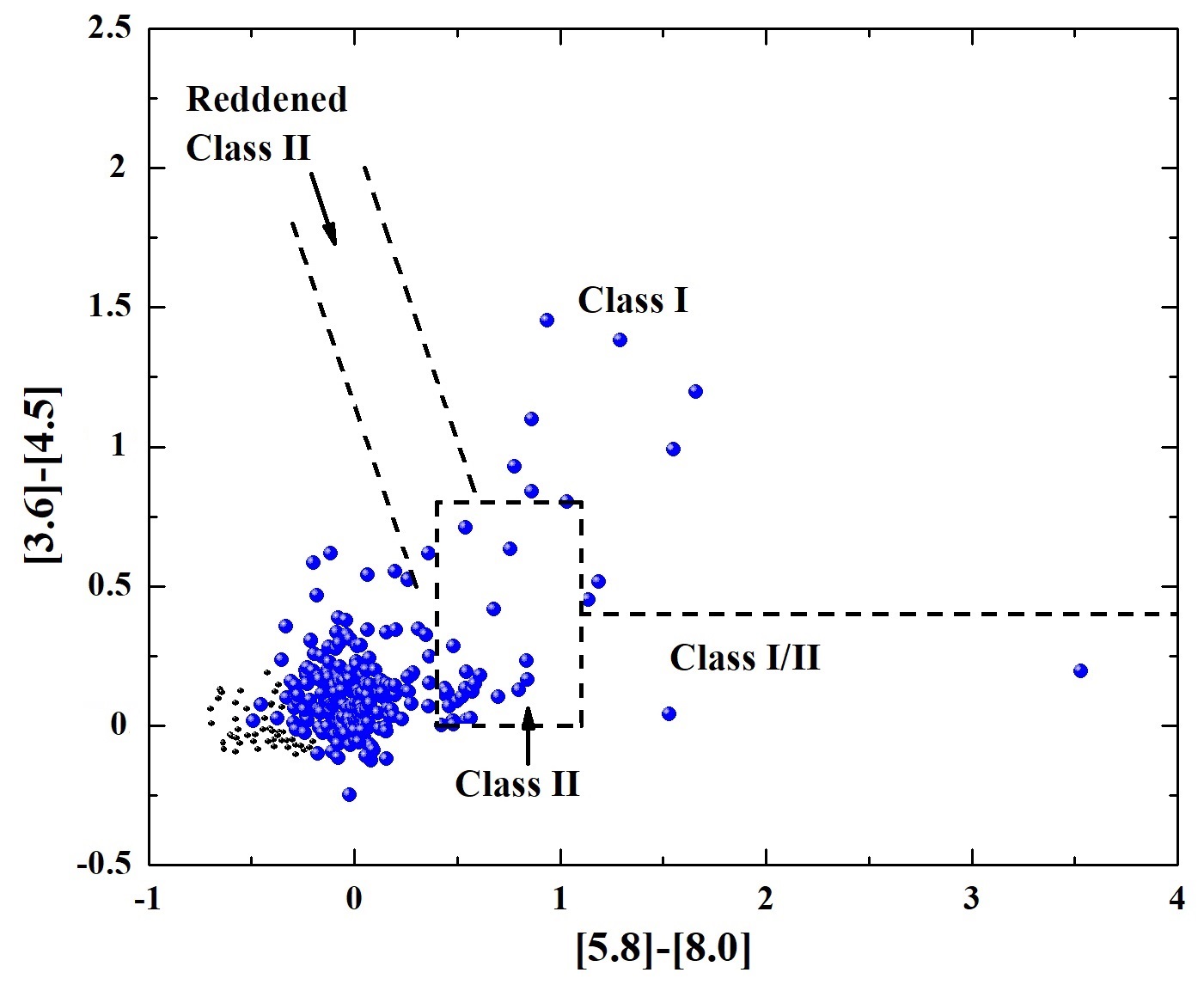}
\includegraphics[width=0.415\linewidth]{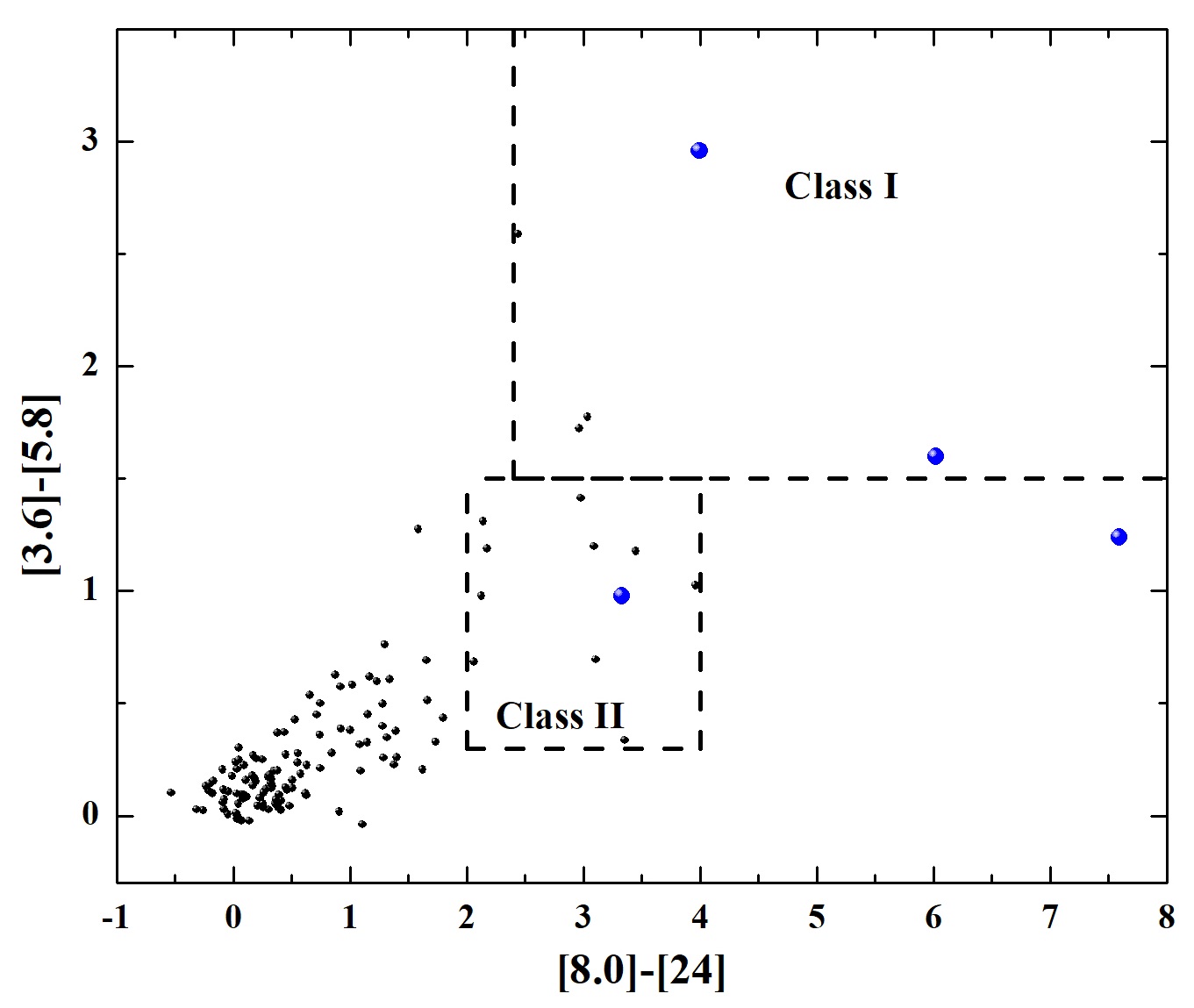}
\includegraphics[width=0.42\linewidth]{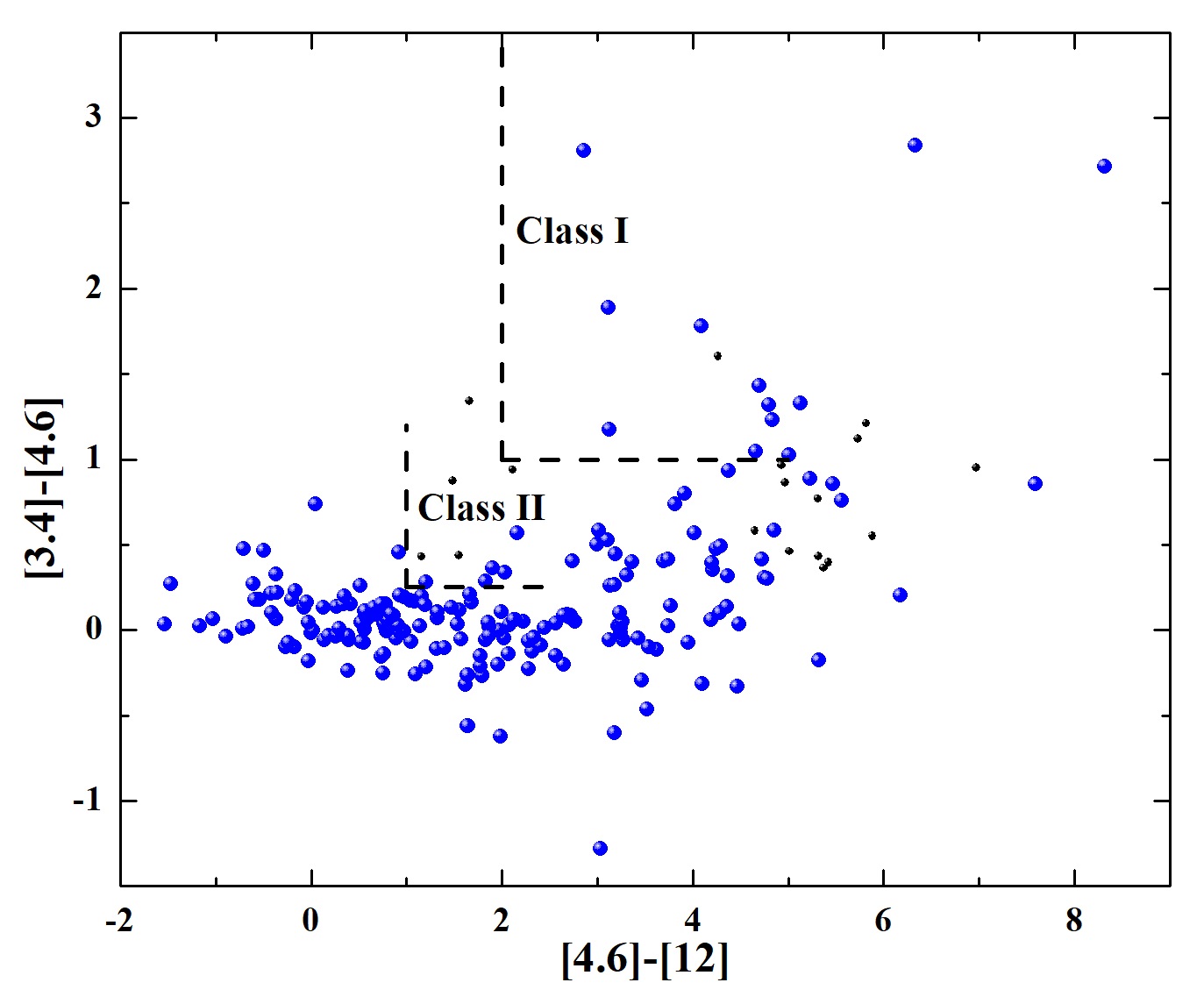}
\includegraphics[width=0.41\linewidth]{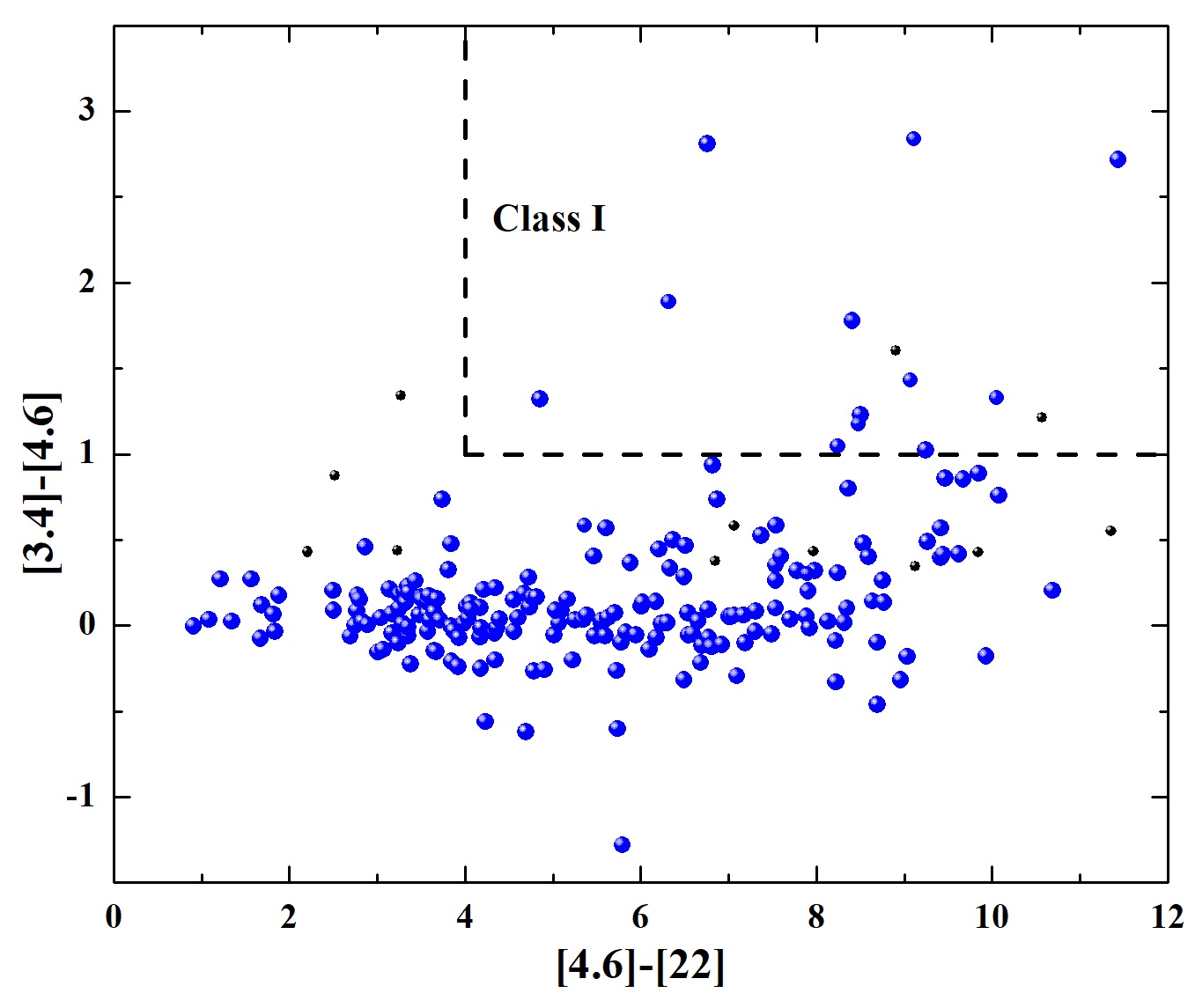}
\caption{Colour-colour diagrams of the region. {\itshape Top left panel}: (J-H) vs. (H-K) diagram. The dwarf and giant loci (solid and dashed curves, respectively) are from \citet{bessell88} and were converted to the CIT system \citep{carpenter01}. The parallel lines represent the interstellar reddening vectors \citep{rieke85}. The locus of unreddened classical T Tauri stars is from \citet{meyer97}. The region bounded by dashed lines is the Herbig Ae/Be stars location \citep{hernandez05}. {\itshape Top right panel}: K-[3.6] vs. [3.6]-[4.5] diagram. In this diagram Class\,I and II domains are separated by the dashed line. The arrow shows the extinction vector \citep{flaherty07}. All the lines are from \citet{allen07}. {\itshape Middle left panel}: [3.6]-[4.5] vs. [5.8]-[8.0] diagram. Two parallel dashed lines border the positions of reddened Class\,II objects. The horizontal dashed line shows the adopted division between Class\,I and Class\,I/II sources. The dashed rectangle shows the position of Class\,II sources. All the lines are from \citet{megeath04}. {\itshape Middle right panel}: [3.6]-[5.8] vs. [8.0]-[24] diagram. The horizontal and vertical dashed lines separate Class\,I sources in this region. The dashed rectangle shows the position of Class\,II sources. All the lines are from \citet{muzerolle04}. {\itshape Bottom left and right panels}: [3.4]-[4.6] vs. [4.6]-[12] and [3.4]-[4.6] vs. [4.6]-[22] diagrams. The blue circles are selected YSO candidates and black circles are non-classified ones. Not all non-classified objects are presented in these diagrams. IRAS\,19111+1048 source is indicated by a red triangle.}
\label{fig:cc}
\end{figure*}

When selecting potential members of the cluster from stars located in the direction of the molecular cloud, we assumed that most of the members of the considered active star-forming region are YSOs. One of the main observational characteristics of YSOs is an IR excess due to the presence of circumstellar discs and envelopes \citep{lada03,hartmann09}; furthermore, the measure of the IR excess in the NIR and/or MIR ranges can be used to characterise the evolutionary stage of a YSO (Class\,I and Class\,II). Therefore, YSO candidates can be identified based on their position in colour--colour (c-c) diagrams. The choice of colours depends on the available data.  We constructed six (c-c) diagrams, as shown in Figure \ref{fig:cc}. The same approach has previously been successfully applied to the IRAS\,05168+3634 star-forming region \citep{azatyan19}.

Objects with IR excess were first identified using the (J-H) versus (H-K) c-c diagram shown in Figure \ref{fig:cc} (top left panel). The solid and dashed curves represent the loci of the intrinsic colours of dwarf and giant stars \citep{bessell88} converted to the CIT system \citep{carpenter01}; the parallel solid lines drawn from the base and tip of these loci are the interstellar reddening vectors \citep{rieke85}. The locus of unreddened classical T Tauri stars (CTTSs) is taken from \citet{meyer97}. The region where the intermediate-mass PMS stars, i.e. Herbig Ae/Be stars, are usually found is bounded by dashed lines \citep{hernandez05}. Objects with different evolutionary stages occupy specific areas of this diagram \citep{lada92}: (i) Classical Be stars, (ii) objects located to the left, and (iii) objects located to the right of reddening vectors. Classical Be stars have J-K\,<\,0.6 and H-K\,<\,0.3 colour indices, which we removed from the final list.

The deviation of YSOs from the main sequence (MS) on this diagram can have two causes: the presence of IR excess and interstellar absorption, which also leads to objects reddening. In the latter case, deviation from the MS will be directed along the reddening vectors. The IR excess of objects located to the right of reddening vectors cannot be caused solely by interstellar absorption and, at least partially, their IR excess is caused by the existence of a circumstellar disc and envelope. Therefore, objects located to the right of the reddening vectors can be considered YSO candidates. Among the objects located in the reddening band of MS and giants, we classified those that have a (J-K)\,>\,3\,mag colour index as Class\,I evolutionary stage YSOs \citep{lada92}. These are located in the upper right corner of the diagram. A total of 5,135 YSO candidates were selected using the (J-H) versus (H-K) c-c diagram.

Other objects in the reddening band are generally considered to be either field stars or Class\,III objects with small NIR excess. The latter objects can be potential members of the star-forming region, however, differentiating between field stars and Class\,III objects is very difficult. Thus, we added to the final list those objects from this sample that were classified as Class\,I and Class\,II evolutionary stage YSOs in at least two other c-c diagrams.

Not all objects in the main sample were detected in the J, H, and K bands simultaneously. Accordingly, we used data from the GLIMPSE catalogue to combine NIR and MIR photometry to identify sources with IR excesses and compile a more complete excess/disc census for the region. Since the 4.5\,$\mu$m band is the most sensitive of the four IRAC bands to YSOs \citep{gutermuth08}, we used a K-[3.6] versus [3.6]-[4.5] c-c diagram. The [3.6]-[4.5] colour is dominated by accretion rate, with higher accretion rate sources appearing redder \citep{allen04}. Figure \ref{fig:cc} (top right panel) shows the K-[3.6] versus [3.6]-[4.5] c-c diagram, with diagonal lines outlining the YSO location region and a dashed line separating the Class\,I and Class\,II object domains. The arrow shows the extinction vector \citep{flaherty07}. All lines are taken from \citet{allen07}. A total of 1,016 YSO candidates were selected using the K-[3.6] versus [3.6]-[4.5] c-c diagram.

The MIR SEDs of Class\,I and Class\,II objects are dominated by emission from dusty circumstellar material, allowing them to be readily distinguished from pure photospheric sources such as unrelated field stars and indistinguishable diskless members. {\itshape Spitzer} is advantageous for studying such regions since its imaging instruments (IRAC at 3.6--8\,$\mu$m and MIPS at 24--160\,$\mu$m) are targeted at the MIR wavelengths, which are less affected by extinction from dust in comparison to NIR \citep{gutermuth08}. Since the IRAC 8\,$\mu$m band overlaps with silicate features, reddened photospheres would appear increasingly blue in [5.8]-[8.0] colour \citep{megeath04}. Figure \ref{fig:cc} (middle left and right panels) shows c-c diagrams with different combinations of {\itshape Spitzer} wavelengths distinguishing YSOs from both Class\,I and Class\,II evolutionary stages. In the middle-left panel, we used {\itshape Spitzer} IRAC wavelengths. Objects located around the ([3.6]-[4.5]; [5.8]-[8.0])=(0;0) region are stars with the colours of the stellar photosphere and diskless PMS (i.e. Class\,III) objects. Class\,II objects fall within the 0\,<\,[3.6]-[4.5]\,<\,0.8 and 0.4\,<\,[5.8]-[8.0]\,<\,1.1 range. Objects with [3.6]-[4.5]\,>\,0.8 and [5.8]-[8.0]\,>\,1.1 range likely correspond to a Class\,I evolutionary stage \citep{allen04,megeath04,qiu08}. Three sources exhibit colours in-consistent with Class\,I, Class\,II, or reddened photosphere models. One source exhibits a higher [3.6]-[4.5] colour value than Class\,II objects, but lower [5.8]-[8.0] colour than Class\,I objects. We identify this source as a reddened Class\,II object. The other two sources have [5.8]-[8.0]\,>\,1.1, consistent with Class\,I objects, but [3.6]-[4.5]\,<\,0.4, which is lower than predicted by Class I models \citep{allen04}. Since these objects share the properties of Class\,I and Class\,II sources, we refer to them as Class\,I/II \citep{megeath04}. In total, 42 YSO candidates were selected using the [3.6]-[4.5] versus [5.8]-[8.0] c-c diagram. 

The 24\,$\mu$m channel is the primary means of identifying optically thin dust discs. Since photospheric colours should be close to zero for all spectral types, the [8.0]–[24] colour is particularly sensitive to excess \citep{muzerolle04}. In the middle right panel of Figure \ref{fig:cc}, we used {\itshape Spitzer} IRAC and MIPS wavelengths to plot a [3.6]-[5.8] versus [8.0]-[24] c-c diagram. Sources clustered around [3.6]-[5.8]\,\textasciitilde\,0 and 0$\leq$[8]–[24]$\leq$1 probably represent a mixture of pure photospheres and perhaps some modest 24\,$\mu$m excess. Class\,I and Class\,II evolutionary stage YSOs have strong [8.0]-[24] excesses and moderate-to-strong [3.6]-[5.8] excesses. Therefore, objects with [3.6]-[5.8]\,>\,1.5 and [8.0]-[24]\,>\,2.4 colours are likely to be Class\,I sources with envelopes. Class\,II objects with optically thick discs exhibit [3.6]-[5.8]\,>\,0.3 and [8.0]-[24]\,>\,2 colour values \citep{muzerolle04,caulet08}. Using the [3.6]-[5.8] versus [8.0]-[24] c-c diagram, we selected 17 YSO candidates.

Additionally, we constructed two other MIR c-c diagrams using the list of objects with good WISE detections, i.e. those possessing photometric uncertainty <\,0.2\,mag in WISE bands. The lower left panel shows the [3.4]-[4.6] versus [4.6]-[12] c-c diagram. Similar to the previous cases, objects with different evolutionary stages fall within certain areas of this diagram \citep{koenig12}. Class\,I YSOs are the reddest objects, with [3.4]-[4.6]\,>\,1.0 and [4.6]-[12]\,>\,2.0. Class\,II YSOs are slightly less red objects and have [3.4]-[4.6]-$\sigma$([3.4]-[4.6])\,>\,0.25 and [4.6]-[12]-$\sigma$([4.6]-[12])\,>\,1.0, where $\sigma$(...) indicates the combined photometric error, added in quadrature. A total of 130 YSO candidates were selected using the [3.4]-[4.6] versus [4.6]-[12] c-c diagram.

The accuracy of the previous classification of stars with photometric errors <\,0.2\,mag can be verified using WISE band 4. Previously classified Class\,I sources were re-classified as Class\,II if [4.6]-[22]\,<\,4.0, and the Class\,II stars were returned to the unclassified pool if [3.4]-[12]\,<\,-1.7 $\times$ ([12]-[22]) + 4.3 \citep{koenig12}. There were no incorrect selections in the pre-classified objects in bands 1--3, confirming the results obtained in the [3.4]-[4.6] versus [4.6]-[12] c-c diagram (Figure \ref{fig:cc}, lower right panel). 

\subsubsection{$\alpha_{IRAC}$ slope}
\label{3.2.3}

Examining the IR SEDs of YSOs is one of the most robust methods for identifying the presence of a circumstellar disc. The shape of the SED can distinguish a disc origin for the observed IR excesses from other possible causes. A particularly useful measure of the shape of a SED is its slope, which is defined as $\lambda F_\lambda \varpropto \lambda^{\alpha_{IRAC}}$ \citep{lada87}. We measured the $\alpha_{IRAC}$ slope values for each of the sources detected in all four IRAC bands (3-8\,$\mu$m). Diskless stars (i.e. stellar photospheres) are characterised by $\alpha_{IRAC}$<-2.56 (Class\,III), while evolved discs fall within the -2.56<$\alpha_{IRAC}$<-1.8 range (Class\,II/III). Class\,II evolutionary stage objects have values in the -1.8< $\alpha_{IRAC}$<0 range, while Class\,I evolutionary stage objects are characterised by $\alpha_{IRAC}$>0 \citep{hartmann05,lada06}. The selection of YSOs by $\alpha_{IRAC}$ slope is based on the same fluxes as used in the [3.6]-[4.5] versus [5.8]-[8.0] c-c diagram, however, in this case, 186 objects were added to the YSO candidates.

\subsubsection{Extraction of field contamination}
\label{3.2.4}

While IR excess is a powerful membership diagnostic tool for young and embedded sources, many potential contaminants exhibit IR excess. There are two main classes of extragalactic contaminants that can be misidentified as YSOs \citep{stern05}. One is star-forming galaxies and narrow-line active galactic nuclei (AGNs), which have growing excesses at 5.8 and 8.0\,$\mu$m due to hydrocarbon emission. The other is broad-line AGNs, which have IRAC colours very similar to those of YSOs. \citet{gutermuth08} developed a method based on the Bootes Shallow Survey data to substantially mitigate these contaminants. In this method, hydrocarbon emission sources, including galaxies and narrow-line AGNs, can be eliminated based on their positions in the [3.6]-[5.8] versus [4.5]-[8.0] diagram. The broad-line AGNs, which are typically fainter than YSOs in the {\itshape Spitzer} bands, are identified by their positions in the [4.5] versus [4.5]-[8.0] colour--magnitude diagram.

We firstly checked our list of YSOs according to the conclusions of \citet{stern05}, i.e. galaxies dominated by PAH emission should have MIR colours that occupy a relatively unique area of most IRAC c-c diagrams: [3.6]-[5.8]\,<\,1.5$\times$([4.5]-[8.0]-1)/2, [3.6]-[5.8]\,<\,1.5, and [4.5]-[8.0]\,>\,1. Notably, no sources on our list satisfied these conditions. By initially filtering out PAH sources, we can construct a filter that more closely traces the broad-line AGN distribution and then use the [4.5] versus [4.5]-[8.0] colour-magnitude diagram to flag likely broad-line AGNs according to the conditions of \citet{gutermuth08} and \citet{qiu08}: [4.5]-[8.0]\,>\,0.5, [4.5]\,>\,13.5+([4.5]-[8.0]-2.3)/0.4, and [4.5]\,>\,13.5. In total, 26 sources satisfied these conditions on our list, 14 of which were already classified as YSOs from other c-c diagrams. Among our selected YSOs, we also applied a condition (i.e. [4.5]\,>\,7.8\,mag and [8.0]-[24.0]\,<\,2.5\,mag) to identify possible AGB contaminants \citep{robitaille08}. In total, 113 sources were identified as possible AGB stars among our list, 64 of which were already classified as YSOs in other c-c diagrams. Thus, in total, we removed 78 objects from the list of YSO candidates.

\subsubsection{YSO candidates}
\label{3.2.5}

The investigated region is quite distant and large, thus, there is a high probability of selecting objects that do not belong to the molecular cloud. On the other hand, PMS objects with comparatively small NIR excesses may be located in the reddening band and are therefore excluded from the selection. To minimise the likelihood of making an incorrect selection, we selected YSOs on the criterion of being stars classified as objects with IR excess by at least two different selection methods, i.e. the c-c diagrams and $\alpha_{IRAC}$ slope. However, since the region has two saturated areas in the MIR band around the IRAS objects (IRAS\,19110+1045 with 25" radius and IRAS\,19111+1048 with 50" radius), this can lead to the potential loss of objects belonging to the molecular cloud. Accordingly, objects within those areas classified as YSOs based on only the NIR c-c diagram were included in the list of candidate YSOs. We selected the dominant evolutionary stage for each object in our chosen methods. In the case of equality, the older stage was selected. In the c-c diagrams, Class\,I and Class\,II YSOs are indicated with blue filled circles. Non-classified objects are shown with black circles.

In total, we selected 909 YSOs within a 6\,arcmin radius, which include 849 Class\,II and 60 Class\,I objects. Among these, 56 objects were selected based on MIR photometric data.

\subsection{SED analysis}
\label{3.3}

To confirm the selected YSOs and to determine their parameters, we constructed their SEDs and fitted them with the radiative transfer models of \citet{robitaille07}. These models assume an accretion scenario in the star formation process, where a central star is surrounded by an accretion disc, an infalling flattened envelope, and the presence of bipolar cavities. We used the command-line version of the SED fitting tool where numerous precomputed models are available. This procedure was performed using wavelengths ranging from 1.1\,$\mu$m to 500\,$\mu$m, in particular, J, H, and K (UKIDSS); 3.6, 4.5, 5.8, and 8.0\,$\mu$m ({\itshape Spitzer} IRAC); 24\,$\mu$m ({\itshape Spitzer} MIPS); 3.4, 4.6, 12, and 22\,$\mu$m (WISE); 70 and 160\,$\mu$m ({\itshape Herschel} PACS); and 70, 160, 250, 350, and 500\,$\mu$m (Hi-GAL). We considered as an upper limit the FIR fluxes that were identified as closest by coordinates to the NIR object to avoid possible errors (see Section \ref{3.2.1}). For the interstellar extinction, we chose an interval of 10–100\,mag that would exceed the results obtained by COBE/DIRBE and IRAS/ISSA maps \citep[A$_v$\,=\,10\,--\,50\,mag,][]{schlegel98}. The distance interval corresponds to the estimates made in the previous studies (6.5–9.5\,kpc, see Introduction).

To identify the representative values of different physical parameters, the tool retrieved the best-fit model and all models that exhibited differences between $\chi^2$ values and the best $\chi^2$ smaller than 3N, where N is the number of data points used \citep{robitaille07}. This approach was chosen because the sampling of the model grid is too sparse to effectively determine the minima of the $\chi^2$ surface and consequently obtain the confidence intervals.

As we selected YSOs in two MIR--saturated regions using only their J, H, and K magnitudes (see Section \ref{3.2.5}), constructing their SEDs based on only three photometric data points does not provide a reliable basis for any conclusions (115 YSOs). Excluding these objects, we achieved relatively robust parameters for 431 of the 793 selected YSOs (16 Class\,I and 415 Class\,II) with $\chi^2$<100 that composes 55\% of the total number. The fact that reliable parameters could not be determined for the remaining 45\% of objects can be explained by several reasons, which likely include errors in the identification and selection of objects.

\begin{figure}[hbt!]
\centering
\includegraphics[width=1\linewidth]{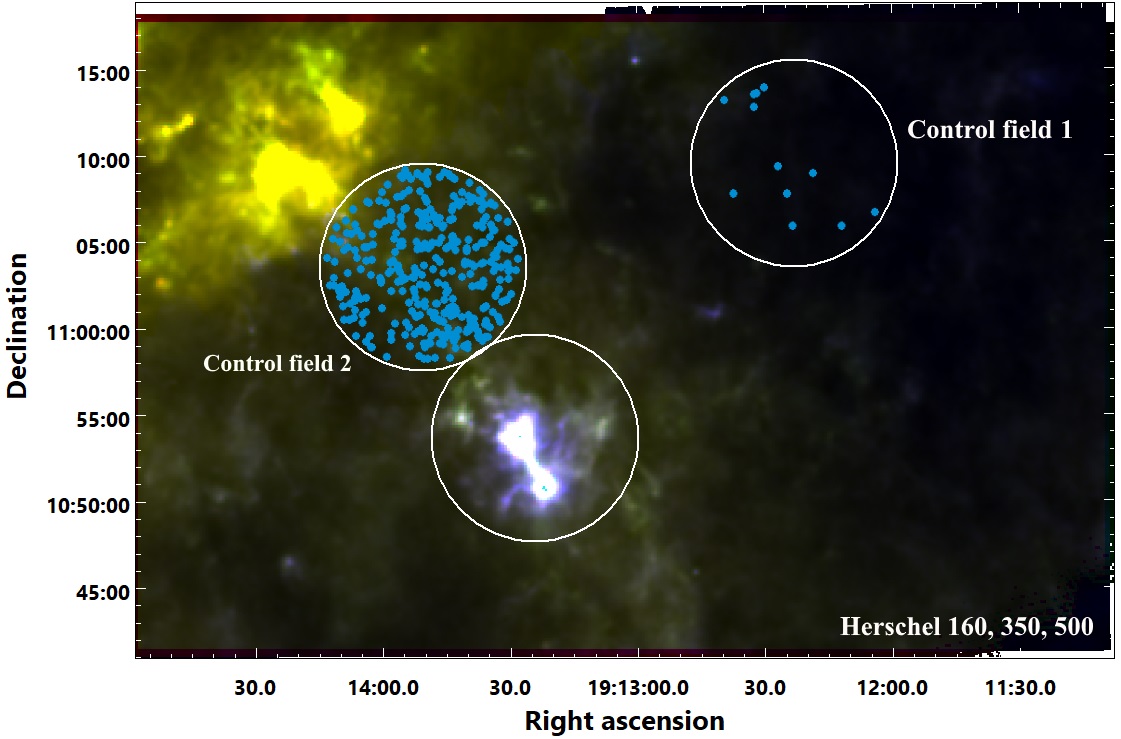}
\caption{{\itshape Herschel}\,160 (blue), 350 (green), 500 (red)\,$\mu$m colour-composite image of the considered region and two Control fields with the same 6\,arcmin radius around the centres with coordinates $\alpha$\,(2000)\,=\,19:12:22.86, $\delta$\,(2000)\,=\,11:09:30.7 (Control field\,1) and $\alpha$\,(2000)\,=\,19:13:52.96, $\delta$\,(2000)\,=\,11:03:23.5 (Control field\,2). YSOs selected by SED fitting tool (see Section \ref{3.3}) in both Control fields are indicated in filled blue circles.}
\label{fig:field}
\end{figure}

To estimate the reliability of the results obtained by the SED fitting tool, we carried out the same analysis in a control field. For Control field\,1, we selected a region located in the vicinity of but outside the GRSMC 45.46+0.05 star-forming region \citep{simon01}. Figure \ref{fig:field} shows the position of Control field\,1 with the same 6\,arcmin radius. The stellar population analysis of the Control field\,1 involved the same steps previously described in Section \ref{3.2}. We selected 250 probable YSO candidates out of 33,676 objects identified in UKIDSS\,DR6. For the SED analysis, we used the same A$_v$ ranges and distances. As a result, we obtained relatively robust parameters for \textit{only} 12 objects with $\chi^2$<100.  

\begin{figure}[hbt!]
\centering
\includegraphics[width=0.9\linewidth]{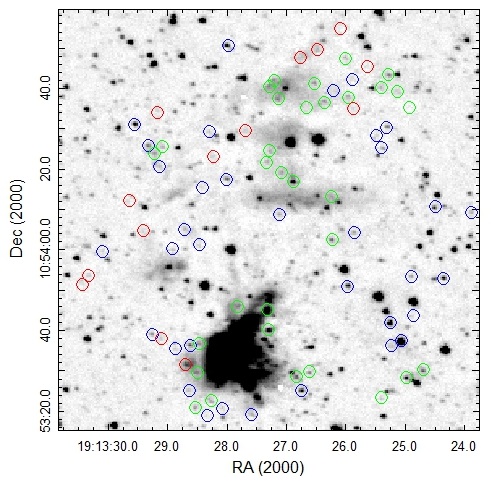}
\includegraphics[width=0.9\linewidth]{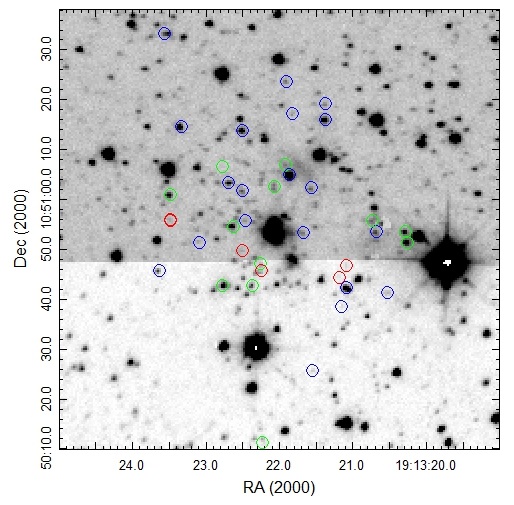}
\caption{K image of G45.07+0.13 (top panel) and G45.12+0.13 (lower panel). The selected YSOs with mergedClass\,=\,-1/-2 (star/probable star marked by blue circles, mergedClass\,=\,+1/-3 (galaxy/probable galaxy) - by green circles. Red circles show the identified non-stellar objects.}
\label{fig:nonst}
\end{figure}

\begin{figure}[hbt!]
\centering
\includegraphics[width=0.8\linewidth]{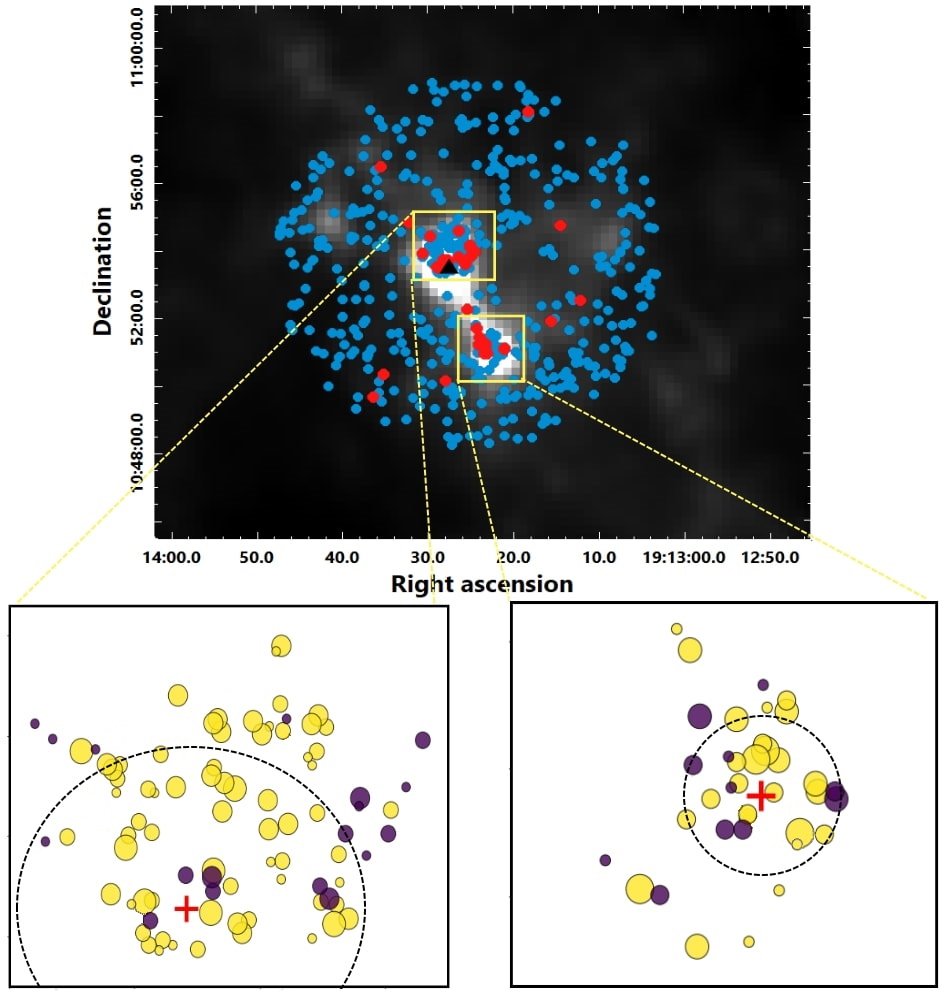}
\caption{\textit{(Top panel)}: Distribution of YSOs in the region on {\itshape Herschel} 500\,$\mu$m image. Class\,I and Class\,II objects are indicated by filled red and blue circles, respectively. IRAS\,19111+1048 source is indicated by a black triangle. \textit{(Bottom panels)}: The two insets show the distribution of the clusters' members. Yellow and purple circles correspond to an older and younger population, respectively. The size of each circle is related to its mass falling within certain interval of masses: 1--3\,M$_\odot$ (smallest), 3--5\,M$_\odot$, 5--7\,M$_\odot$, >7\,M$_\odot$ (largest). The colour and size of the members are taken based on their positions on the CMD (Figure \ref{fig:CMD} left panel). Red crosses show the coordinates of IRASs. Dashed circles shows the MIR-saturated regions around the IRAS sources.}
\label{fig:YSOs}
\end{figure}

The stellar content analysis of Control field\,1 allows us to conclude that the 431 YSOs ($\chi^2$<100) selected by the SED fitting tool in our considered region are located with high probability at the same distance as two UC\,HII regions. By examining the distances corresponding to the models obtained with the SED fitting tool, we found that, for most objects, the best models correspond to a distance of 7.8\,kpc. For the other 45\% of the objects, the SED fitting tool did not yield a reliable result. In addition to the aforementioned reasons, this issue may relate to the presence of fore/background objects. The latter issue is especially likely because the region is distant.

\subsection{Final catalogue}
\label{3.7}

As noted earlier, we included objects in the initial UKIDSS\,DR6 selection with the catalogue parameter mergedClass\,=\,+1 (galaxy) and -3 (probable galaxy). Among these, there may be some artefacts or so-called non-stellar objects \citep{Solin2012}. The reasons for their formation may be manifold, including nebulous structures, diffraction patterns of bright stars, bright stars at or near the border of the detector array, etc. To justify the reliability of the final list, we reviewed our sample for artefacts. For this purpose, following the advice of previous works \citep[e.g. ][]{lucas08,Solin2012}, the parameter k\_1ppErrBits was used. This parameter contains the quality error information for each source detected by the K filter. Among the 431 objects with robust SEDs, only one showed k\_1ppErrBits = 64 (i.e. bad pixel(s)) and seven showed k\_1ppErrBits = 4194304 (i.e. lies within a dither offset of the stacked frame boundary); we thus excluded these from the final list. Such a small percentage of artefacts ($\sim$\,2\%) indicates sufficiently reliable filtering of stellar objects using the SED fitting tool.

We also performed a visual inspection of the 115 YSO candidates that were selected based only on their NIR photometric data. From our point of view, these objects are of the greatest interest as they are located in the immediate vicinity of the UC\,HIIs. The analysis results are shown in Figure \ref{fig:nonst}. The surface brightness distribution of 20 faint objects indicates that they can be attributed to extended/non-stellar objects rather than point sources. In UKIDSS DR6, they were identified as galaxies/probable galaxies (mergedClass\,=\,+1/-3). These objects are marked with red circles in Figure \ref{fig:nonst}. To ensure the purity of the sample, we removed these objects from the final list.

Overall, the final list comprised 518 YSOs (423 with constructed SEDs and 95 YSOs in two saturated regions). The coordinates, NIR and MIR photometric data, $\alpha_{IRAC}$, and evolutionary stage of the 518 YSOs are presented in Table \ref{tab:NIR}, while Table \ref{tab:FIR} presents the FIR fluxes of the selected objects. In total, we identified 23 objects in FIR bands, including the object associated with IRAS\,19111+1048. Table \ref{tab:SED} shows the weighted means and the standard deviation values of parameters for all models with $\chi^2-\chi^2_{best}$<3N obtained by the SED fitting tool. In Tables \ref{tab:NIR}, \ref{tab:FIR}, and \ref{tab:SED}, the parameters of the members of the region are listed in the following order: first, the IRAS\,19111+1048 source then objects within the IRAS\,19111+1048 and IRAS\,19110+1045 clusters (see Section \ref{3.4}), and, finally, those within a 6\,arcmin radius of the whole region that are not included in the clusters. The numbering of objects in Table \ref{tab:SED} was performed according to Tables \ref{tab:NIR} and \ref{tab:FIR}. Table \ref{tab:SED} contains the parameters of the 423 objects for which the SED fitting tool achieved a value of $\chi^2$<100.

Some conclusions can be drawn based on the data obtained by the SED fitting tool. The average interstellar extinction value is equal to A$_v$=13\,mag, which corresponds to a lower interstellar extinction estimate than that obtained from the  COBE/DIRBE and IRAS/ISSA maps \citep{schlegel98}. The average mass of the YSOs is approximately 4.4\,M$_\odot$, with a minimum estimated mass of 1.7\,M$_\odot$ and a maximum of 22\,M$_\odot$. Primarily, the lack of low-mass stellar objects can be explained by the large distance of the star-forming region.

\subsection{Distribution of YSOs}
\label{3.4}

\begin{figure*}[hbt!]
\centering
\includegraphics[width=0.5\linewidth]{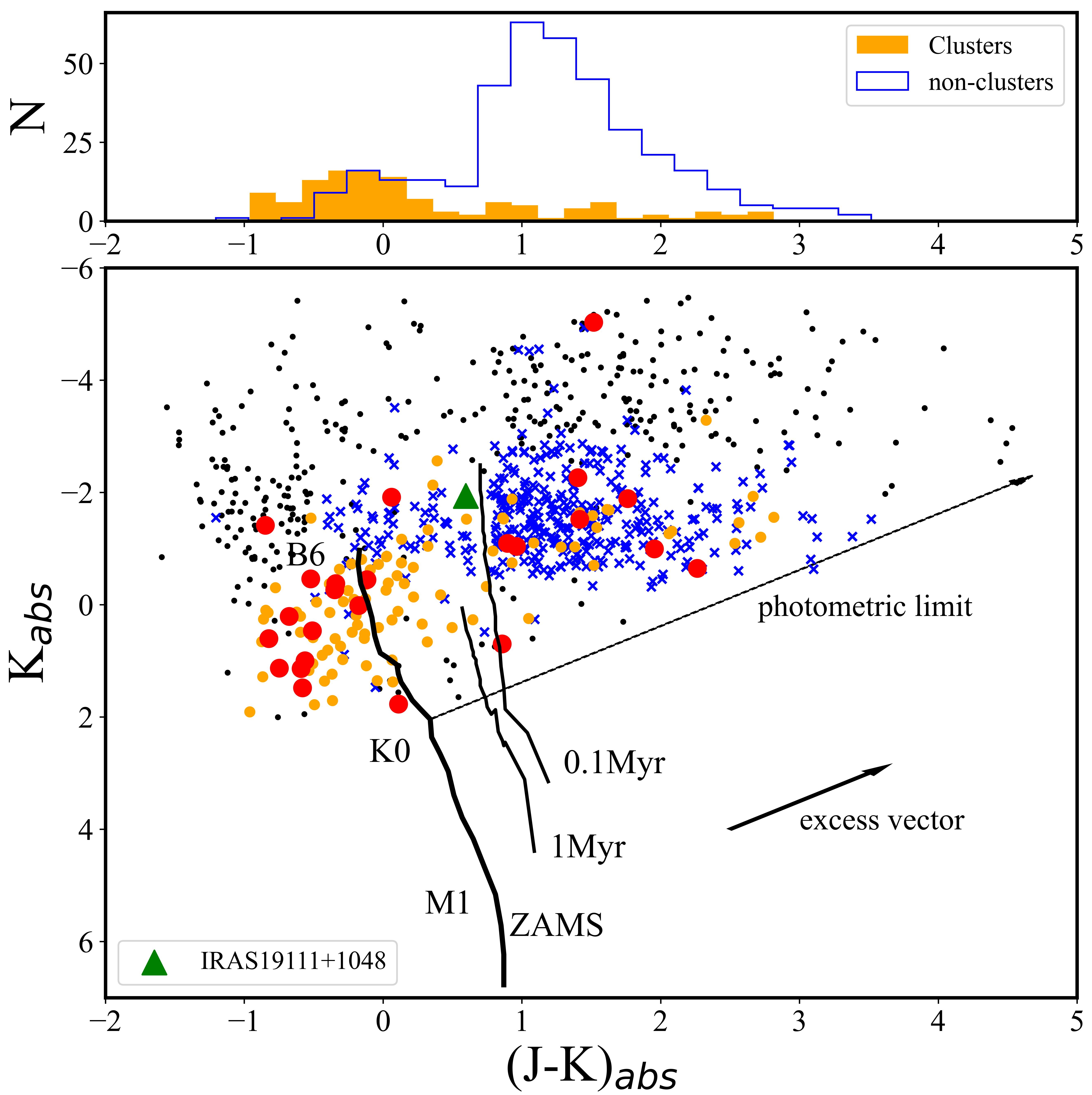}
\includegraphics[width=0.48\linewidth]{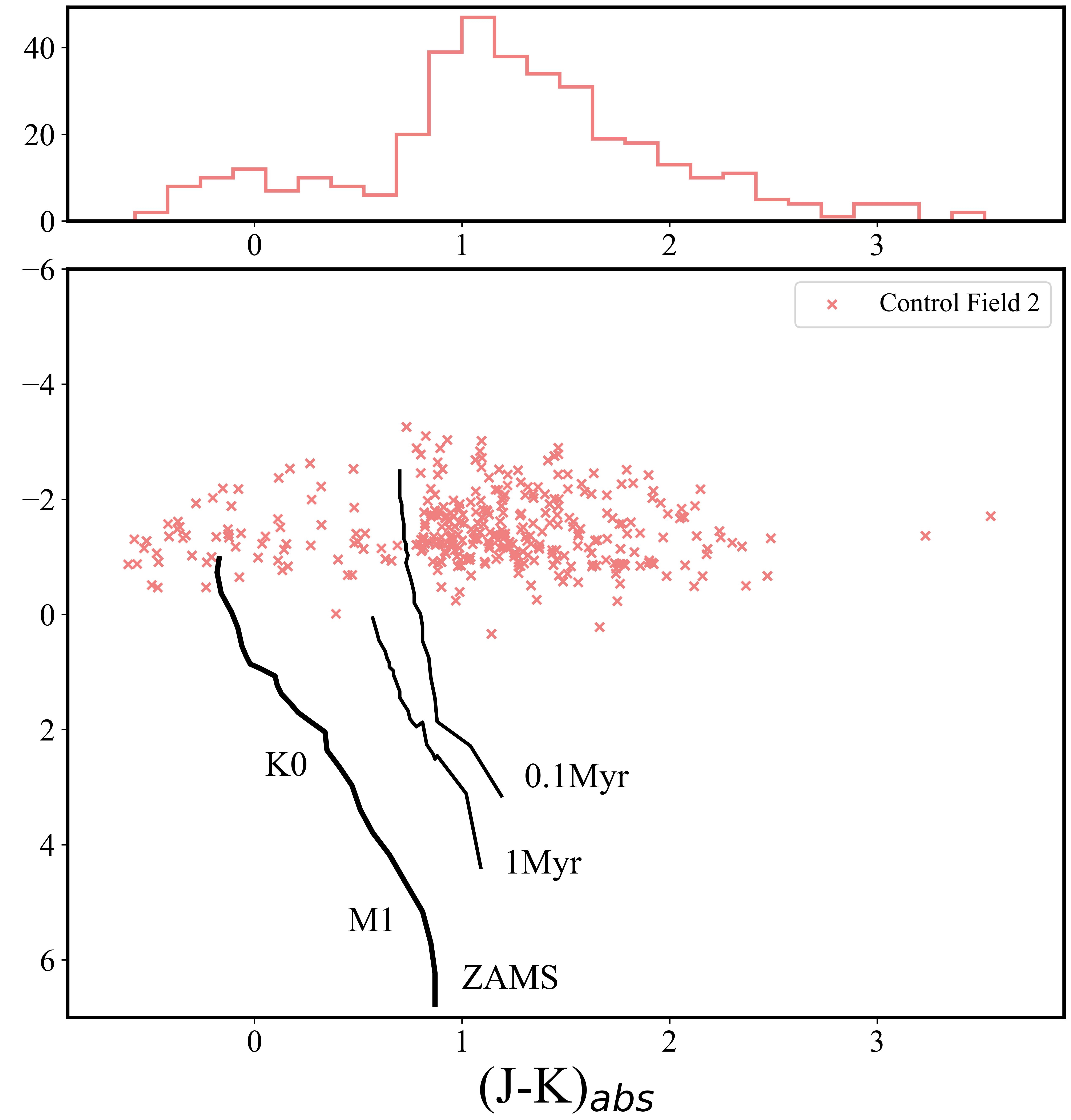}
\caption{K vs. (J–K) colour-magnitude diagrams for identified YSOs in the considered region \textit{(bottom left panel)} and Control field\,2 \textit{(bottom right panel)}. The PMS isochrones for the 0.1 and 1\,Myr \citep{siess00} and ZAMS are drawn as solid thin and thick lines, respectively. The positions of a few spectral types are labelled. The J and K magnitudes of the YSOs are corrected for the interstellar extinctions determined according to average A$_v$\,=\,13\,mag value obtained by the SED fitting tool. \textit{(bottom left panel)}: Red circles are stellar objects within the IRAS clusters with constructed SED based on more than 3 photometric data points. Objects located in the saturated regions around two IRAS sources are yellow circles. Non-cluster objects are blue crosses and no-SED objects are black dots. IRAS\,19111+1048 source is indicated by a green triangle and labelled. The solid arrow indicates the average slope of NIR excesses caused by circumstellar discs \citep{lopez07}. The dashed arrow indicates the photometric limit of UKIDSS in K-band. \textit{(bottom right panel)}: Stellar objects located in Control field\,2 are indicated by coral crosses. \textit{(Top left and right panels):} Histograms of (J-K)$_{abs}$ values.}
\label{fig:CMD}
\end{figure*}

The top panel of Fig.~\ref{fig:YSOs} shows the distribution of the selected YSOs in the field, with Class\,I and Class\,II objects shown by filled red and blue circles, respectively. Excluding the regions in the vicinity of the IRAS sources, all types of stellar objects are distributed relatively homogeneously in the molecular cloud. Additionally, in both UC\,HII regions, close to the IRAS sources, the selected YSOs form relatively dense concentrations or clusters. These concentrations were revealed above by the radial distribution of stellar densities (see Section \ref{3.2.1} and Figure \ref{fig:rVSd}). The existence of the cluster around IRAS\,19111+1048 is more obvious now. We refined the radius of each cluster relative to its geometric centre based on the density distribution of the selected YSOs. The stellar density was determined for each ring of width 0.1\,\'\, by dividing the number of YSOs by the surface area. The radius of each cluster was considered the distance at which fluctuations in the rings' stellar density became random according to Poisson statistics. Table \ref{tab:IRAS} presents the coordinates of the geometric centres of the clusters in the vicinity of IRAS sources and the whole region in Cols. 2 and 3, the coordinates of IRAS sources in Cols. 4 and 5, and the radii of the clusters in Col. 6. Cols. 7 and 8 present the stellar content and surface density of each cluster, as well as of the whole region, based on our selection. The surface density of stars in clusters is four times higher than that of the entire region. The radii of the clusters are in good agreement with the sizes of the UC\,HII regions as determined by the distributions of dust temperature and column density (see Section \ref{3.1}). Note that in previous studies, the young massive (OB) population embedded in the innermost regions of the clump has already been reported  \citep[e.g.][]{vig06,Rivera2010}. The two insets in the lower panels of Figure~\ref{fig:YSOs} show the distribution of the members in the dense clusters. The yellow and purple circles correspond to older and younger populations, respectively. The size of each circle is related to its mass falling within certain mass intervals: 1--3\,M$_\odot$ (smallest), 3--5\,M$_\odot$, 5--7\,M$_\odot$, and >7\,M$_\odot$ (largest). The colour and size of the members are taken based on their positions on the CMD (see Figure \ref{fig:CMD} left panel). The surface density distribution of the YSOs did not show any concentration around the BGPS\,6737 (MSX G045.1663+00.0910) object.

\citet{vig06} proposed that the initial trigger and power source of G45.12+0.13 is the brightest radio source S14, which was deduced to be of spectral type O6 from integrated radio emission. They also found two NIR objects, IR4 and IR5, within S14. According to our data, the nearest object to the S14 peak ($\sim$\,7.7") that can satisfy the conditions defined in \citet{vig06} is a star with 9.4\,$\pm$\,4.3\,M$_{\odot}$ mass, 23,000\,$\pm$\,11,000\,K temperature, and (2.5\,$\pm$\,1.2)\,$\times$\,10$^6$ years evolutionary age. This star is mentioned as IRAS 19111+1048 in Tables \ref{tab:NIR}, \ref{tab:FIR}, \ref{tab:SED}, and the NIR c-c diagram. It is also the closest object ($\sim$\,9") to the brightest source A16-24 obtained by the CORNISH interferometer at 4.8\,GHz \citep{Rivera2010}. This stellar object has II evolutionary class. Besides IRAS\,19111+1048 source, two selected YSOs are also located within the radio contours of S14, i.e. stars ID\,52 and 53 in Table \ref{tab:NIR}, which are located at distances about 8.5 and 11" from the S14 peak, respectively. This group of objects can be assumed to be the power source of G45.12+0.13. \citet{liu19} also suggested that S14 is a protocluster hosting several ZAMS stars.

\begin{table}
\caption{Comparison of IR and radio data}
\resizebox{1\textwidth}{!}{
\label{tab:vig}
\begin{tabular}{lcc|lcc} \toprule
ID & GMRT & CORNISH & ID & GMRT & CORNISH \\
\hline\noalign{\smallskip}
(1) & (2) &  (3) & (1) & (2) & (3) \\
\hline \noalign{\smallskip}
10 & S3 & $-$ & 63, 68 & S18 & $-$\\
11 & S6 & $-$ & 64 & $-$ & A16$-$C32\\
16 & S5 & $-$ & 65, 71 & S16 & $-$\\
26, 36 & S10 & $-$ & 66 & $-$ & A16$-$C33\\
37 & S9 & $-$ & 74 & S19 & $-$\\
43, 45 & S12 & $-$ & 87, 88 & S24 & $-$\\
50, 53 & $-$ & A16$-$C8 & 91 & S25 & $-$\\
56 & $-$ & A16$-$C25 & 98 & S27 & A12$-$C1\\
58 & S15 & $-$ & 120 & $-$ & A16$-$C5  \\
59, 60 & $-$ & A16$-$C27 & 105 & $-$ & A12$-$C3\\
\bottomrule
\end{tabular}
}
\textit{\textbf{Notes.}} (1) - ID of objects from Table \textit{NIR}, (2) - sources obtained by radio continuum observations at 1280, 610\,MHz frequency bands \citep{vig06}, (3)- sources obtained by CORNISH interferometer at 4.8 GHz \citep{Rivera2010}. 

\end{table}

According to \citet{vig06}, IRAS 19110+1045 is associated with the compact radio source S27, which has no NIR counterparts. Using the SED fitting tool, \citet{persi19} obtained embedded pair of YSOs with the mass of 24\,M$_{\odot}$ associated with IRAS\,19110+1045. Unfortunately, due to the saturation of the central parts of UC\,HII regions in the MIR range, we were unable to identify the YSOs associated with this source.

We examined the location of identified stellar objects in relation to radio sources \citep{vig06,Rivera2010}. Table \ref{tab:vig} includes objects that are located no further than 5" from the peak of the radio sources. For the majority, we see that at least one stellar object can be associated with a radio source from \citet{vig06}.

\begin{table}
\caption{Properties of the region}
\resizebox{1\textwidth}{!}{
\label{tab:IRAS}
\begin{tabular}{l c cccc c c} \toprule
Name & $\alpha$(2000) & $\delta$(2000) & $\alpha$(2000) & $\delta$(2000) & Radius & N & Density\\
& (hh mm ss) & (dd mm ss) & (hh mm ss) & (dd mm ss) & (arcmin) & & (arcmin$^{-2}$)\\
\hline\noalign{\smallskip}
(1) & (2) &  (3) & (4) & (5) & (6) & (7) & (8)\\
\hline \noalign{\smallskip}
IRAS\,19111+1048 & 19 13 27.8 & +10 53 36.74 & 19 13 27.3 & +10 54 07.42 & 1.2 & 87 & 19.2\\
IRAS\,19110+1045 & 19 13 22.0 & +10 50 54.00 & 19 13 23.1 & +10 50 51.88 & 0.8 & 37 & 18.4\\
Region & $-$ & $-$ & 19 13 24.2 & +10 53 38.00 & 6 & 518 & 4.6\\ 
\bottomrule
\end{tabular}
}
\textit{\textbf{Notes.}} (1)-names of (sub-)regions, (2),(3)- coordinates of the IRAS sources, (4),(5)- coordinates of the geometric centres, (6)-the radii of (sub-)regions, (7)- numbers of objects within the selected radii, (8)-surface stellar density in the (sub-)regions.

\end{table}

\subsection{Colour-magnitude diagram}
\label{3.5}

A colour-magnitude diagram (CMD) is a useful tool to study the nature of the stellar population within star-forming regions. The K versus J–K diagram provides the maximum contrast between the IR excess produced by the presence of a disc, which mostly affects K, and interstellar extinction, which has a greater effect on J. The distribution of the identified YSOs in the K versus J–K CMD is shown in the left panel of Fig. \ref{fig:CMD}. The ZAMS (thick solid curve) and PMS isochrones for the 0.1 and 1\,Myr ages (thin solid curves) are taken from \citet{siess00}. We used the conversion table from \citet{kenyon94}. Circles indicate stars in the IRAS clusters, while red circles are stars with more than three photometric measurements. For these stars, the SED fitting tool obtained reliable parameters (i.e. $\chi^2$\,<\,100). Yellow circles represent objects in the saturated regions that have only three photometric measurements. Objects located outside of the two IRAS clusters (hereafter \textit{non-cluster}) for which the SED fitting tool obtained reliable parameters are marked by blue crosses. The black dots are objects with no reliable parameters ($\chi^2$\,>\,100). To correct the J and K magnitudes of the selected YSOs, we used a 7.8\,kpc distance and the average interstellar extinction value (A$_v$\,=\,13\,mag, see Section \ref{3.3}). NIR excess, shown in the c-c diagrams (Fig. \ref{fig:cc}), is usually caused by the presence of disks around young stars; thus, by incorporating theoretical accreting disk models, the excess effect on the CMD can be accurately represented by vectors of approximately constant slope for disks around Class\,II T Tauri stars. The components of the vector are (1.01, -1.105) and (1.676, 1.1613) in magnitude units \citep{lopez07}. More massive YSOs are usually much more embedded than T Tauri stars, thus, this correction is unlikely to apply to such objects. However, the presence of a spherical envelope around the disc should cause a greater decrease in J-K for the same variation in the K than in the case of a ``naked'' disc \citep{cesaroni15}; accordingly, the \citet{lopez07} correction can be used to obtain a lower limit of stellar mass in the region. Using the excess vector, we also determined the photometric limit of the UKIDSS data represented by the dashed arrow in the diagram; this is parallel to the excess vector and passes through the ZAMS point with coordinates (0.34, 2.05). The Y-coordinate corresponds to the photometric limit of UKIDSS in the K band (18.02 mag) corrected for distance and interstellar extinction. The photometric limit corresponds to the ZAMS stars with 1.4\,M$_{\odot}$ mass; this value is in good agreement with the minimum mass (i.e. 1.7\,M$_{\odot}$), obtained by the SED fitting tool. We observe that all detected objects are brighter in the K-band than the photometric limit of the UKIDSS survey. Undoubtedly, the large distance and interstellar extinction play a crucial role in this result.

Several conclusions can be drawn based on the location of stellar objects in the CMD. Objects without reliable parameters (black dots) occupy a separate region; in general, these are brighter in the K-band than objects with reliable parameters and most of them are likely foreground objects. The stars that belong to the star-forming region (circles and crosses) have considerable variation concerning their colour index, i.e. (J-K)$_{abs}$. The positions of objects in the two IRAS clusters and non-cluster regions are different. An overwhelming majority (more than 80\%) of the non-cluster objects are located to the right of the 0.1\,Myr isochrone. In contrast, about 75\% of objects in the IRAS clusters are located to the left of the 0.1\,Myr isochrone and concentrated around the ZAMS, some of which fall to the left of the ZAMS. For improved clarity, the histograms of (J-K)$_{abs}$ are shown in the top panels of Fig. \ref{fig:CMD}. In general, the evolutionary age spread of the vast majority of stellar objects from both samples is small; furthermore, the members of the IRAS clusters are more developed than the non-cluster objects.

\begin{figure}
\centering
\includegraphics[width=0.8\linewidth]{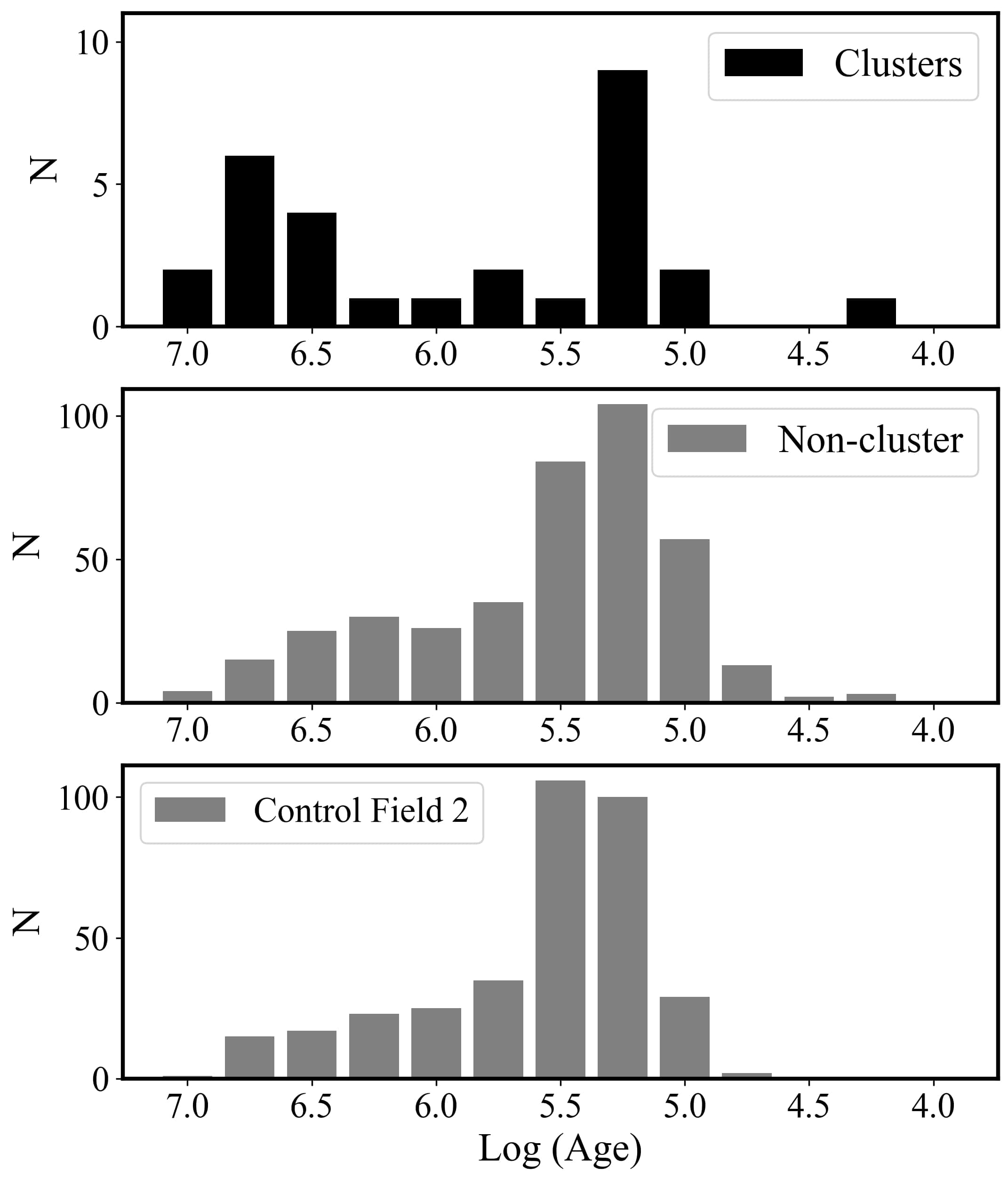}
\caption{Histogram of evolutionary ages (by the SED fitting tool) for members of the IRAS clusters (top panel), the non-cluster objects (middle panel), and the objects in the Control field\,2 (bottom panel). The bin size corresponds to Log\,(Age)\,=\,0.25.}
\label{fig:Age}
\end{figure}

The middle panel of Figure \ref{fig:Age} shows the distribution of evolutionary ages (by the SED fitting tool) for the non-cluster objects, which confirms the above interpretation. The age distribution of these objects has a well-defined peak at Log(Age)\,$\approx$\,5.25. In contrast, the distribution of the evolutionary ages of the objects in the clusters has two peaks (Fig. \ref{fig:Age} top panel) centred at Log(Age)\,$\approx$\,6.75 and Log(Age)\,$\approx$\,5.25. Note that the age distribution of the objects in the clusters was constructed based on parameters from only 29 YSOs for which the SED fitting tool was applied. Most of the other 95 YSOs in the MIR-saturated regions are concentrated around the ZAMS and to the left of the 1\,Myr isochrone. Therefore, we assume that these objects will have a real contribution to the first peak in the evolutionary age distribution. Accordingly, this distribution will have only one well-defined peak as per the histogram of (J-K)$_{abs}$) around Log(Age)\,$\approx$\,6.75. 

The cluster identification confirms the assumptions made in previous studies based mainly on radio observations of star clusters near the IRAS sources, which include high- and intermediate-mass ZAMS objects \citep[e.g.][and ref. therein]{liu19,Rivera2010,vig06}. The clusters' members exhibit low scatter relative to the isochrones. The clusters' origin can be assumed to relate to an external triggering shock. We examined the distribution of the cluster members relative to their masses and evolutionary ages (see the lower panels of Fig. \ref{fig:YSOs}), which demonstrated that the objects are not segregated according to these parameters. The members of the IRAS\,19111+1048 cluster are asymmetrically arranged relative to the IRAS source, and, like the isotherms (see Fig. \ref{fig:3}), are elongated in a northwest direction. There are potentially two UC\,HIIs herein (i.e. G45.12+0.13 and G45.13+0.14) that are separated from each other in a northwest direction \citep{Fuente2020a}.

The non-cluster objects of all evolutionary classes are uniformly distributed across the field (see Fig. \ref{fig:YSOs} top panel). As shown in previous studies, several massive stellar objects in or near the ZAMS are responsible for the ionisation of UC\,HII regions and can trigger a second star formation event along the ionising radiation ridge \citep[e.g.][]{Blum2008, Paron2009}. Accordingly, the newly formed stars are unevenly distributed and localised in the radiation ridge. In our study, the non-cluster YSOs are found to be uniformly distributed in the molecular cloud. Therefore, the origin of the non-cluster objects cannot be explained by the activity of the embedded massive stars in the UC\,HII regions. We instead assume that these uniformly distributed objects are part of the young stellar population of the GRSMC\,45.46+0.05 molecular cloud, which is an active star-forming region \citep{simon01}.

To confirm this assumption, we performed the same analysis in Control field\,2, an area we selected that is located in the GRSMC\,45.46+0.05 star-forming region and very close to the considered region. Fig. \ref{fig:field} shows the position of Control field\,2 with the same 6\,arcmin radius. The analysis of the stellar population of Control field\,2 includes the same steps described in Section \ref{3.2}. We selected 777 YSO candidates out of 30,254 objects identified in UKIDSS\,DR6. For the SED analysis of these YSO candidates, we used the same A$_v$ and distance ranges of 10–100\,mag and 6.5–9.5\,kpc, respectively. We obtained relatively robust parameters for $\sim$350 objects ($\chi^2$<100); therefore, the number of YSOs in Control field\,2 is almost the same as the number of non-cluster objects in the considered region. The mass range of these objects also coincides (from 1.9 to 18\,M$_{\odot}$). The evolutionary age distributions of the non-cluster stellar objects and Control field\,2 objects show significant similarity (see bottom panel of Fig. \ref{fig:Age}), with the Control field\,2 object ages exhibiting one well-defined peak. The distribution of evolutionary ages and the peak value (Log(Age)\,$\approx$\,5.35) coincide with those of the non-cluster objects. To further confirm the similarity between the non-cluster and Control field\,2 stellar objects, we also used CMD. The distribution of the objects from Control field\,2 in the K versus J–K CMD is shown in the right panel of Fig. \ref{fig:CMD} (coral crosses). We used the same correction as previously for the J and K photometry, i.e. 7.8\,kpc distance and A$_v$\,=\,13\,mag interstellar extinction. More than 80\% of the objects are located to the right of the 0.1\,Myr isochrone, like the non-cluster objects. Thus, the main parameters (evolutionary ages, masses, and surface stellar density) of the non-cluster and Control field\,2 objects are almost the same. Accordingly, our assumption that the non-cluster YSOs are part of the young stellar population of the GRSMC\,45.46+0.05 molecular cloud is plausible. To understand the tracers of their origins, the star formation history of the GRSMC\,45.46+0.05 star-forming region as a whole must be investigated.

\begin{figure}
\centering
\includegraphics[width=0.9\linewidth]{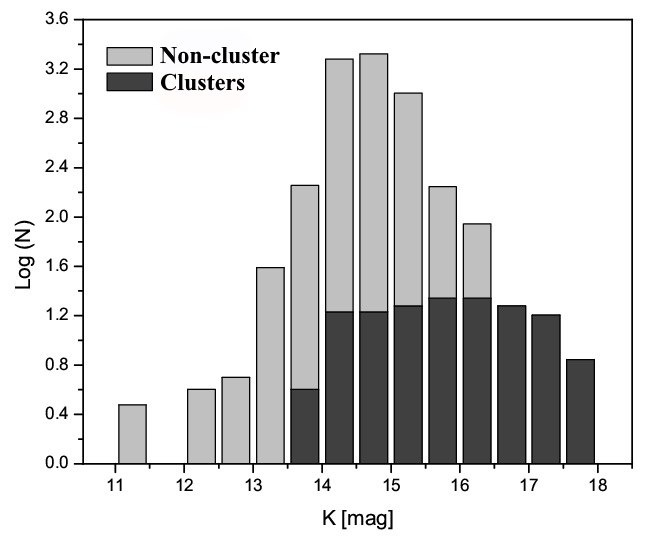}
\caption{K luminosity functions derived for the IRAS clusters (dark grey) and non-cluster objects (light grey) as histograms of the number of stars in logarithm vs. apparent K magnitude. The bin size corresponds to 0.5\,mag.}
\label{fig:KLF}
\end{figure}

\subsection{K luminosity function and Initial mass function}
\label{3.6}

The luminosity function in the K-band is frequently used in studies of young clusters and star-forming regions as a diagnostic tool of their IMF and the star formation history of their stellar populations \citep{zinnecker93, lada95, jose12}. The observed KLF is a result of the IMF, PMS evolution, and star-forming history; thus, the KLF slope is a potential age indicator of young clusters \citep{lada96}. The KLF can be defined as $dN(K)/dK \propto 10^{\alpha K}$, where $\alpha$ is the slope of the power-law and N(K) is the number of stars brighter than K\,mag. We evaluated the $\alpha$ slopes individually for the stellar populations of the IRAS clusters and the non-cluster objects. The KLF slopes were estimated by fitting the number of YSOs in 0.5\,mag bins using a linear least-squares fitting routine. Since the interstellar extinction is independent of stellar mass, it does not affect the result \citep{megeath96}; therefore we can use all the sources in the sample to define the slopes without the complication of extinction correction. Figure \ref{fig:KLF} shows the observed overplotted KLFs of the IRAS clusters (dark grey) and the non-cluster objects (light grey). No sharp decline is recorded at the end of KLF as a whole, i.e. it occurs gradually, showing that the photometric completeness limit of the survey did not strongly affect the shape of the KLF. Note that the KLF of the IRAS clusters does not show an obvious maximum while the KLF of the non-cluster objects has a maximum at the same magnitude as the whole region. This is most likely due to the quantitative ratio between the objects from different samples. Its slope is also noticeably steeper than that of the IRAS clusters' KLF. Thus, the IRAS clusters appear to show unique characteristics relative to the whole region, once again confirming that they are at different stages of evolution. The calculated KLF $\alpha$ slopes for the IRAS clusters and non-cluster objects were 0.23\,$\pm$\,0.10 and 0.55\,$\pm$\,0.09, respectively. 

One of the most fundamental disciplines of astrophysical research is the origin of stars and stellar masses. The IMF, together with the star formation rate, dictates the evolution and fate of both galaxies and star clusters \citep{kroupa02}. The study of the IMFs of star-forming regions is important -- their mass functions (MFs) can be considered as IMFs because they are too young to lose a significant number of members through dynamical or stellar evolution effects. We assume that the MF and the mass-luminosity relation for a stellar cluster are described by power laws like the KLF, i.e. they are of the form $dN(logm)\propto m^{\gamma}$ and $L_K \propto m^{\beta}$, where $\gamma$ and $\beta$ are the slopes of MF and mass-luminosity relation, respectively. We obtained two completely different KLF results for the two IRAS clusters and the non-cluster objects. Here, we discuss the IMF of each case separately as the initial conditions may differ. The mass-luminosity relation with $\beta$\,\textasciitilde\,1 is mostly used for stars at the lower mass end of the IMF, i.e. G–M stars for the 1\,Myr cluster \citep{lada93b, megeath96}. If $\beta$\,\textasciitilde\,1, values such as $\alpha=\gamma/(2.5\beta)=0.68$ and 0.54 would be expected for power-law IMFs with slopes $\gamma=1.7$ \citep{miller79, scalo86} and 1.35 \citep{salpeter55}, respectively. The latter value is in good agreement with our KLF fit result for the non-cluster stellar objects (0.55\,$\pm$\,0.09). If we assume $\beta$\,\textasciitilde\,2, typically used for a larger and higher mass range (O–F stars) at 1\,Myr \citep{balog04}, this results in $\alpha=0.27$ for a Salpeter-type IMF and $\alpha=0.34$ for a Miller-Scalo type IMF. The Salpeter-type IMF is closer to the KLF fit result for the two IRAS clusters (0.23\,$\pm$\,0.10). The results obtained via the IMFs are well-correlated with the masses obtained from the SED fitting tool (see Section \ref{3.3}). Notably, the 1\,Myr age estimate for the non-cluster objects does not match with the results of the SED fitting tool and CMD (see Figures \ref{fig:Age} and \ref{fig:CMD}) -- in general, the non-cluster objects are noticeably younger. Such a discrepancy is because the mass-luminosity relation with $\beta\sim$\,1 is used mostly for low-mass stars, while the SED fitting tool did not find any low-mass objects in the region and the CMD shows that the photometric limit of UKIDSS in the K-band influenced the detection of faint objects in the region. Thus, the resulting loss of low-mass YSOs in the region will strongly affect the results of the star formation rate and star formation efficiency calculation, so we did not define these parameters.

\section{Summary and conclusions}
\label{5}

An infrared study of the star-forming region in a molecular cloud, which includes the UC\,HII regions G45.12+0.13/ IRAS\,19111+1048 and G45.07+0.13/IRAS\,19110+1045, was undertaken with two major components: (i) determination of the ISM physical parameters (i.e. N(H$_2$) and T$_d$) and (ii) searching for and studying the young stellar population.

To determine N(H$_2$) and T$_d$, we applied modified blackbody fitting on {\itshape Herschel} images obtained in four bands: 160, 250, 350, and 500\,$\mu$m. The maps of these parameters allowed us to conclude:

\begin{itemize}
    \item  Within the G45.07+0.13 and G45.12+0.13 regions, N(H$_2$) varies from $\sim$\,3.0\,$\times$\,10$^{23}$ to 5.5\,$\times$\,10$^{23}$\,cm$^{-2}$.
	\item The T$_d$ maximum in G45.12+0.13 is 35\,K and in G45.07+0.13 it is 42\,K.
	\item The dust temperature drops significantly from the centre to the periphery, reaching values of around 18--20 K. The T$_d$ decline levels out at distances of $\sim$\,2.6\,pc and $\sim$\,3.7\,pc from IRAS\,19110+1045 and IRAS\,19111+1048, respectively. The masses of the gas-dust matter are $\sim$\,1.7\,$\times$\,10$^5\,M_\odot$ and $\sim$\,3.4\,$\times$\,10$^5\,M_\odot$ in the G45.07+0.13 and G45.12+0.13 regions, respectively.
	\item The column density map revealed a bridge between the two UC\,HII regions with relatively high density (N(H$_2$) =\,4.3$\times$10$^{23}$\,cm$^{-2}$) and low temperature (T$_d$\,=\,19\,K), which is also clearly visible on {\itshape Herschel} images. This suggests that these two UCHII regions are physically connected.
	\end{itemize}

The objectives of studying the stellar population were to identify the members of the clusters, associated with the UC\,HII regions, as well as to determine their main parameters (e.g. stellar masses, evolutionary ages, and age spread). The identification and classification of YSOs using NIR, MIR, and FIR photometric data were based on one of the main properties of young stars, namely, their infrared excess due to the presence of circumstellar discs and envelopes. The SED fitting tool \citep{robitaille06} was used to determine their main parameters. Within a 6\,arcmin radius around the UC\,HIIs, we obtained relatively robust parameters for 423 YSOs (16 Class\,I and 407 Class\,II objects) with $\chi^2$\,<\,100. We also identified 95 YSOs located in the two MIR-saturated regions based only on the J, H, and K photometric data. A detailed study of stellar objects in the considered star-forming region made it possible to obtain the following results:

\begin{itemize}
    \item The stellar density radial distribution shows the existence of dense clusters in the vicinity of both IRAS sources. These clusters include 37 and 87 members in G45.07+0.13 and G45.12+0.13, respectively. Their surface stellar density exceeds the average over the star-forming region around fourfold. The radii are in good agreement with the sizes of the UC\,HII regions as determined by the distributions of dust temperature and column density. The remaining 394 non-cluster objects, irrespective of their evolutionary classes, are uniformly distributed in the molecular cloud.
    \item We were unable to identify stellar objects with masses less than 1.4\,M$_\odot$. This can be explained by the large distance of the star-forming region.
    \item The study of the stellar parameters from different samples (i.e. clusters and non-cluster) showed differences between the two populations.
    \item On the CMD, around 75\% of the YSOs in the IRAS clusters are located to the left of the 0.1\,Myr isochrone and are concentrated around the ZAMS. The slope $\alpha$ of the KLF for these objects is 0.23\,$\pm$\,0.10, which agrees well with a Salpeter-type IMF ($\gamma$=1.35) for a high mass range (O–F stars, $\beta\sim$\,2) at 1\,Myr.
    \item The detailed study of the clusters made it possible to identify the stellar object associated with IRAS\,19111 +1048, and this can be the exciting star of G45.12+0.13. This is a star with 9.4\,$\pm$\,4.3\,M$_{\odot}$ mass, 23,000\,$\pm$\,11,000\,K temperature, and (2.5\,$\pm$\,1.2)\,$\times$\,10$^6$\,yr evolutionary age.
    \item The median value of the evolutionary ages obtained by the SED fitting tool for the non-cluster objects is Log(Age)\,$\approx$\,5.25, more than 80\% of which are located to the right of the 0.1\,Myr isochrone. The slope $\alpha$ of the KLF for these objects is 0.55\,$\pm$\,0.09, which agrees better with a Salpeter-type IMF for low-mass objects (G–M stars, $\beta\sim$\,1).
\end{itemize}

Based on the results, we concluded that dense clusters were formed in both UC\,HII regions, which include high- and intermediate-mass stellar objects. The evolutionary ages of these stars, in most cases, are several million years. Likely, low-mass stellar objects were not identified due to the large distance of the star-forming region. The small spread of evolutionary ages suggests that the clusters owe their origin to a triggering shock. Presumably, the low-density extended emission observed on the MIR images \citep{Fuente2020a}, which also stands out well on the dust temperature maps in both UC\,HII regions, may be due to the existence of the stellar clusters. 

The distribution of the non-cluster objects in the molecular cloud implies that their origin cannot be explained by the activity of the embedded massive star(s) in the UC\,HII regions. We assume that these uniformly distributed objects are part of the young stellar population of the GRSMC 45.46+0.05 molecular cloud, which is an active star-forming region. To understand the tracers of their origins, it is necessary to investigate the star formation history of the GRSMC 45.46+0.05 star-forming region as a whole. This extended star-forming region, spreading over an area of about 6\,deg$^2$ in Aquila, is known to host numerous UC\,HII regions, maser emissions, and outflows from young and highly embedded OB stellar clusters \citep[e.g.][]{simon01,Rivera2010}. This work is a part of the project to study the UC\,HII regions in GRSMC\,45.46+0.05, and, consequently, massive star formation and its influence on the surrounding ISM.

\begin{acknowledgement}
We thank the anonymous reviewer for constructive comments and suggestions. We thank Lex Kaper for carefully reading the manuscript and for his helpful remarks. This work was made possible by a research grant number №\,21AG-1C044 from Science Committee of Ministry of Education, Science, Culture and Sports RA. This research has made use of the data obtained at UKIRT, which is supported by NASA and operated under an agreement among the University of Hawaii, the University of Arizona, and Lockheed Martin Advanced Technology Center; operations are enabled through the cooperation of the East Asian Observatory. We gratefully acknowledge the use of data from the NASA/IPAC Infrared Science Archive, which is operated by the Jet Propulsion Laboratory, California Institute of Technology, under contract with the National Aeronautics and Space Administration. We thank our colleagues in the GLIMPSE and MIPSGAL {\itshape Spitzer} Legacy Surveys. This publication also made use of data products from {\itshape Herschel} ESA space observatory. 
\end{acknowledgement}


\bibliography{example}

\begin{thebibliography}{}
\expandafter\ifx\csname natexlab\endcsname\relax\def\natexlab#1{#1}\fi

\bibitem[{{Allen} {et~al.}(2007){Allen}, {Megeath}, {Gutermuth}, {Myers},
  {Wolk}, {Adams}, {Muzerolle}, {Young}, \& {Pipher}}]{allen07}
{Allen}, L., {Megeath}, S.~T., {Gutermuth}, R., {et~al.} 2007, Protostars and
  Planets V, 361

\bibitem[{{Allen} {et~al.}(2004){Allen}, {Calvet}, {D'Alessio}, {Merin},
  {Hartmann}, {Megeath}, {Gutermuth}, {Muzerolle}, {Pipher}, {Myers}, \&
  {Fazio}}]{allen04}
{Allen}, L.~E., {Calvet}, N., {D'Alessio}, P., {et~al.} 2004, \apjs, 154, 363

\bibitem[{{Argon} {et~al.}(2000){Argon}, {Reid}, \& {Menten}}]{argon00}
{Argon}, A.~L., {Reid}, M.~J., \& {Menten}, K.~M. 2000, \apjs, 129, 159

\bibitem[{{Azatyan}(2019)}]{azatyan19}
{Azatyan}, N.~M. 2019, \aap, 622, A38

\bibitem[{{Azatyan} {et~al.}(2016){Azatyan}, {Nikoghosyan}, \&
  {Khachatryan}}]{azatyan16}
{Azatyan}, N.~M., {Nikoghosyan}, E.~H., \& {Khachatryan}, K.~G. 2016,
  Astrophysics, 59, 339

\bibitem[{{Balog} {et~al.}(2004){Balog}, {Kenyon}, {Lada}, {Barsony},
  {Vink{\'o}}, \& {G{\'a}spa{\'r}}}]{balog04}
{Balog}, Z., {Kenyon}, S.~J., {Lada}, E.~A., {et~al.} 2004, \aj, 128, 2942

\bibitem[{{Battersby} {et~al.}(2011){Battersby}, {Bally}, {Ginsburg},
  {Bernard}, {Brunt}, {Fuller}, {Martin}, {Molinari}, {Mottram}, {Peretto},
  {Testi}, \& {Thompson}}]{battersby11}
{Battersby}, C., {Bally}, J., {Ginsburg}, A., {et~al.} 2011, \aap, 535, A128

\bibitem[{{Bessell} \& {Brett}(1988)}]{bessell88}
{Bessell}, M.~S., \& {Brett}, J.~M. 1988, \pasp, 100, 1134

\bibitem[{{Blum} \& {McGregor}(2008)}]{Blum2008}
{Blum}, R.~D., \& {McGregor}, P.~J. 2008, \aj, 135, 1708

\bibitem[{{Breen} {et~al.}(2019){Breen}, {Contreras}, {Dawson}, {Ellingsen},
  {Voronkov}, \& {McCarthy}}]{breen19}
{Breen}, S.~L., {Contreras}, Y., {Dawson}, J.~R., {et~al.} 2019, \mnras, 484,
  5072

\bibitem[{{Carey} {et~al.}(2009){Carey}, {Noriega-Crespo}, {Mizuno}, {Shenoy},
  {Paladini}, {Kraemer}, {Price}, {Flagey}, {Ryan}, {Ingalls}, {Kuchar},
  {Pinheiro Gon{\c{c}}alves}, {Indebetouw}, {Billot}, {Marleau}, {Padgett},
  {Rebull}, {Bressert}, {Ali}, {Molinari}, {Martin}, {Berriman}, {Boulanger},
  {Latter}, {Miville-Deschenes}, {Shipman}, \& {Testi}}]{carey09}
{Carey}, S.~J., {Noriega-Crespo}, A., {Mizuno}, D.~R., {et~al.} 2009, \pasp,
  121, 76

\bibitem[{{Carpenter}(2001)}]{carpenter01}
{Carpenter}, J.~M. 2001, \aj, 121, 2851

\bibitem[{{Caulet} {et~al.}(2008){Caulet}, {Gruendl}, \& {Chu}}]{caulet08}
{Caulet}, A., {Gruendl}, R.~A., \& {Chu}, Y.~H. 2008, \apj, 678, 200

\bibitem[{{Cesaroni} {et~al.}(2015){Cesaroni}, {Massi}, {Arcidiacono},
  {Beltr{\'a}n}, {Persi}, {Tapia}, {Molinari}, {Testi}, {Busoni}, {Riccardi},
  {Boutsia}, {Bisogni}, {McCarthy}, \& {Kulesa}}]{cesaroni15}
{Cesaroni}, R., {Massi}, F., {Arcidiacono}, C., {et~al.} 2015, \aap, 581, A124

\bibitem[{{Churchwell}(2002)}]{churchwell02}
{Churchwell}, E. 2002, \araa, 40, 27

\bibitem[{{Churchwell} {et~al.}(2010){Churchwell}, {Sievers}, \&
  {Thum}}]{churchwell10}
{Churchwell}, E., {Sievers}, A., \& {Thum}, C. 2010, \aap, 513, A9

\bibitem[{{Churchwell} {et~al.}(1992){Churchwell}, {Walmsley}, \&
  {Wood}}]{churchwell92}
{Churchwell}, E., {Walmsley}, C.~M., \& {Wood}, D.~O.~S. 1992, \aap, 253, 541

\bibitem[{{Churchwell} {et~al.}(2009){Churchwell}, {Babler}, {Meade},
  {Whitney}, {Benjamin}, {Indebetouw}, {Cyganowski}, {Robitaille}, {Povich},
  {Watson}, \& {Bracker}}]{churchwell09}
{Churchwell}, E., {Babler}, B.~L., {Meade}, M.~R., {et~al.} 2009, \pasp, 121,
  213

\bibitem[{{de la Fuente} {et~al.}(2020{\natexlab{a}}){de la Fuente}, {Porras},
  {Trinidad}, {Kurtz}, {Kemp}, {Tafoya}, {Franco}, \&
  {Rodr{\'\i}guez-Rico}}]{Fuente2020a}
{de la Fuente}, E., {Porras}, A., {Trinidad}, M.~A., {et~al.}
  2020{\natexlab{a}}, \mnras, 492, 895

\bibitem[{{de la Fuente} {et~al.}(2020{\natexlab{b}}){de la Fuente}, {Tafoya},
  {Trinidad}, {Porras}, {Nigoche-Netro}, {Kemp}, {Kurtz}, {Franco}, \&
  {Rodr{\'\i}guez-Rico}}]{Fuente2020b}
{de la Fuente}, E., {Tafoya}, D., {Trinidad}, M.~A., {et~al.}
  2020{\natexlab{b}}, \mnras, 497, 4436

\bibitem[{{Egan} {et~al.}(2003){Egan}, {Price}, {Kraemer}, {Mizuno}, {Carey},
  {Wright}, {Engelke}, {Cohen}, \& {Gugliotti}}]{egan03}
{Egan}, M.~P., {Price}, S.~D., {Kraemer}, K.~E., {et~al.} 2003, VizieR Online
  Data Catalog, 5114

\bibitem[{{Ellsworth-Bowers} {et~al.}(2015){Ellsworth-Bowers}, {Rosolowsky},
  {Glenn}, {Ginsburg}, {Evans}, {Battersby}, {Shirley}, \&
  {Svoboda}}]{ellsworth15}
{Ellsworth-Bowers}, T.~P., {Rosolowsky}, E., {Glenn}, J., {et~al.} 2015, \apj,
  799, 29

\bibitem[{{Elmegreen} \& {Lada}(1977)}]{elmegreen77}
{Elmegreen}, B.~G., \& {Lada}, C.~J. 1977, \apj, 214, 725

\bibitem[{{Fazio} {et~al.}(2004){Fazio}, {Hora}, {Allen}, {Ashby}, {Barmby},
  {Deutsch}, {Huang}, {Kleiner}, {Marengo}, {Megeath}, {Melnick}, {Pahre},
  {Patten}, {Polizotti}, {Smith}, {Taylor}, {Wang}, {Willner}, {Hoffmann},
  {Pipher}, {Forrest}, {McMurty}, {McCreight}, {McKelvey}, {McMurray}, {Koch},
  {Moseley}, {Arendt}, {Mentzell}, {Marx}, {Losch}, {Mayman}, {Eichhorn},
  {Krebs}, {Jhabvala}, {Gezari}, {Fixsen}, {Flores}, {Shakoorzadeh}, {Jungo},
  {Hakun}, {Workman}, {Karpati}, {Kichak}, {Whitley}, {Mann}, {Tollestrup},
  {Eisenhardt}, {Stern}, {Gorjian}, {Bhattacharya}, {Carey}, {Nelson},
  {Glaccum}, {Lacy}, {Lowrance}, {Laine}, {Reach}, {Stauffer}, {Surace},
  {Wilson}, {Wright}, {Hoffman}, {Domingo}, \& {Cohen}}]{fazio04}
{Fazio}, G.~G., {Hora}, J.~L., {Allen}, L.~E., {et~al.} 2004, \apjs, 154, 10

\bibitem[{{Fish} {et~al.}(2003){Fish}, {Reid}, {Wilner}, \&
  {Churchwell}}]{fish03}
{Fish}, V.~L., {Reid}, M.~J., {Wilner}, D.~J., \& {Churchwell}, E. 2003, \apj,
  587, 701

\bibitem[{{Flaherty} {et~al.}(2007){Flaherty}, {Pipher}, {Megeath}, {Winston},
  {Gutermuth}, {Muzerolle}, {Allen}, \& {Fazio}}]{flaherty07}
{Flaherty}, K.~M., {Pipher}, J.~L., {Megeath}, S.~T., {et~al.} 2007, \apj, 663,
  1069

\bibitem[{{Garay} \& {Lizano}(1999)}]{garay99}
{Garay}, G., \& {Lizano}, S. 1999, \pasp, 111, 1049

\bibitem[{{Griffin} {et~al.}(2010){Griffin}, {Abergel}, {Abreu}, {Ade},
  {Andr{\'e}}, {Augueres}, {Babbedge}, {Bae}, {Baillie}, {Baluteau}, {Barlow},
  {Bendo}, {Benielli}, {Bock}, {Bonhomme}, {Brisbin}, {Brockley-Blatt},
  {Caldwell}, {Cara}, {Castro-Rodriguez}, {Cerulli}, {Chanial}, {Chen},
  {Clark}, {Clements}, {Clerc}, {Coker}, {Communal}, {Conversi}, {Cox},
  {Crumb}, {Cunningham}, {Daly}, {Davis}, {de Antoni}, {Delderfield}, {Devin},
  {di Giorgio}, {Didschuns}, {Dohlen}, {Donati}, {Dowell}, {Dowell}, {Duband},
  {Dumaye}, {Emery}, {Ferlet}, {Ferrand}, {Fontignie}, {Fox}, {Franceschini},
  {Frerking}, {Fulton}, {Garcia}, {Gastaud}, {Gear}, {Glenn}, {Goizel},
  {Griffin}, {Grundy}, {Guest}, {Guillemet}, {Hargrave}, {Harwit}, {Hastings},
  {Hatziminaoglou}, {Herman}, {Hinde}, {Hristov}, {Huang}, {Imhof}, {Isaak},
  {Israelsson}, {Ivison}, {Jennings}, {Kiernan}, {King}, {Lange}, {Latter},
  {Laurent}, {Laurent}, {Leeks}, {Lellouch}, {Levenson}, {Li}, {Li},
  {Lilienthal}, {Lim}, {Liu}, {Lu}, {Madden}, {Mainetti}, {Marliani}, {McKay},
  {Mercier}, {Molinari}, {Morris}, {Moseley}, {Mulder}, {Mur}, {Naylor},
  {Nguyen}, {O'Halloran}, {Oliver}, {Olofsson}, {Olofsson}, {Orfei}, {Page},
  {Pain}, {Panuzzo}, {Papageorgiou}, {Parks}, {Parr-Burman}, {Pearce},
  {Pearson}, {P{\'e}rez-Fournon}, {Pinsard}, {Pisano}, {Podosek}, {Pohlen},
  {Polehampton}, {Pouliquen}, {Rigopoulou}, {Rizzo}, {Roseboom}, {Roussel},
  {Rowan-Robinson}, {Rownd}, {Saraceno}, {Sauvage}, {Savage}, {Savini},
  {Sawyer}, {Scharmberg}, {Schmitt}, {Schneider}, {Schulz}, {Schwartz},
  {Shafer}, {Shupe}, {Sibthorpe}, {Sidher}, {Smith}, {Smith}, {Smith},
  {Spencer}, {Stobie}, {Sudiwala}, {Sukhatme}, {Surace}, {Stevens}, {Swinyard},
  {Trichas}, {Tourette}, {Triou}, {Tseng}, {Tucker}, {Turner}, {Vaccari},
  {Valtchanov}, {Vigroux}, {Virique}, {Voellmer}, {Walker}, {Ward}, {Waskett},
  {Weilert}, {Wesson}, {White}, {Whitehouse}, {Wilson}, {Winter}, {Woodcraft},
  {Wright}, {Xu}, {Zavagno}, {Zemcov}, {Zhang}, \& {Zonca}}]{griffin10}
{Griffin}, M.~J., {Abergel}, A., {Abreu}, A., {et~al.} 2010, \aap, 518, L3

\bibitem[{{Gutermuth} {et~al.}(2008){Gutermuth}, {Myers}, {Megeath}, {Allen},
  {Pipher}, {Muzerolle}, {Porras}, {Winston}, \& {Fazio}}]{gutermuth08}
{Gutermuth}, R.~A., {Myers}, P.~C., {Megeath}, S.~T., {et~al.} 2008, \apj, 674,
  336

\bibitem[{{Han} {et~al.}(2015){Han}, {Zhou}, {Wang}, {Esimbek}, {Zhang}, \&
  {Wang}}]{han15}
{Han}, X.~H., {Zhou}, J.~J., {Wang}, J.~Z., {et~al.} 2015, \aap, 576, A131

\bibitem[{{Hartmann}(2009)}]{hartmann09}
{Hartmann}, L. 2009, {Accretion Processes in Star Formation: Second Edition}
  (Cambridge University Press)

\bibitem[{{Hartmann} {et~al.}(2005){Hartmann}, {Megeath}, {Allen}, {Luhman},
  {Calvet}, {D'Alessio}, {Franco-Hernandez}, \& {Fazio}}]{hartmann05}
{Hartmann}, L., {Megeath}, S.~T., {Allen}, L., {et~al.} 2005, \apj, 629, 881

\bibitem[{{Hern{\'a}ndez} {et~al.}(2005){Hern{\'a}ndez}, {Calvet}, {Hartmann},
  {Brice{\~n}o}, {Sicilia-Aguilar}, \& {Berlind}}]{hernandez05}
{Hern{\'a}ndez}, J., {Calvet}, N., {Hartmann}, L., {et~al.} 2005, \aj, 129, 856

\bibitem[{{Hern{\'a}ndez-Hern{\'a}ndez}
  {et~al.}(2019){Hern{\'a}ndez-Hern{\'a}ndez}, {Kurtz}, {Kalenskii},
  {Golysheva}, {Garay}, {Zapata}, \& {Bergman}}]{hernandez19}
{Hern{\'a}ndez-Hern{\'a}ndez}, V., {Kurtz}, S., {Kalenskii}, S., {et~al.} 2019,
  \aj, 158, 18

\bibitem[{{Hern{\'a}ndez-Hern{\'a}ndez}
  {et~al.}(2014){Hern{\'a}ndez-Hern{\'a}ndez}, {Zapata}, {Kurtz}, \&
  {Garay}}]{hernandez14}
{Hern{\'a}ndez-Hern{\'a}ndez}, V., {Zapata}, L., {Kurtz}, S., \& {Garay}, G.
  2014, \apj, 786, 38

\bibitem[{{Hildebrand}(1983)}]{hildebrand83}
{Hildebrand}, R.~H. 1983, \qjras, 24, 267

\bibitem[{{Hofner} \& {Churchwell}(1996)}]{hofner96}
{Hofner}, P., \& {Churchwell}, E. 1996, \aaps, 120, 283

\bibitem[{{Hunter} {et~al.}(1997){Hunter}, {Phillips}, \& {Menten}}]{hunter97}
{Hunter}, T.~R., {Phillips}, T.~G., \& {Menten}, K.~M. 1997, \apj, 478, 283

\bibitem[{{Jose} {et~al.}(2012){Jose}, {Pandey}, {Ogura}, {Samal}, {Ojha},
  {Bhatt}, {Chauhan}, {Eswaraiah}, {Mito}, {Kobayashi}, \& {Yadav}}]{jose12}
{Jose}, J., {Pandey}, A.~K., {Ogura}, K., {et~al.} 2012, \mnras, 424, 2486

\bibitem[{{Kauffmann} {et~al.}(2008){Kauffmann}, {Bertoldi}, {Bourke}, {Evans},
  \& {Lee}}]{kauffmann2008}
{Kauffmann}, J., {Bertoldi}, F., {Bourke}, T.~L., {Evans}, N.~J., I., \& {Lee},
  C.~W. 2008, \aap, 487, 993

\bibitem[{{Kenyon} {et~al.}(1994){Kenyon}, {Gomez}, {Marzke}, \&
  {Hartmann}}]{kenyon94}
{Kenyon}, S.~J., {Gomez}, M., {Marzke}, R.~O., \& {Hartmann}, L. 1994, \aj,
  108, 251

\bibitem[{{Keto}(2007)}]{keto07}
{Keto}, E. 2007, \apj, 666, 976

\bibitem[{{Koenig} {et~al.}(2012){Koenig}, {Leisawitz}, {Benford}, {Rebull},
  {Padgett}, \& {Assef}}]{koenig12}
{Koenig}, X.~P., {Leisawitz}, D.~T., {Benford}, D.~J., {et~al.} 2012, \apj,
  744, 130

\bibitem[{{K{\"o}nyves} {et~al.}(2015){K{\"o}nyves}, {Andr{\'e}},
  {Men'shchikov}, {Palmeirim}, {Arzoumanian}, {Schneider}, {Roy}, {Didelon},
  {Maury}, {Shimajiri}, {Di Francesco}, {Bontemps}, {Peretto}, {Benedettini},
  {Bernard}, {Elia}, {Griffin}, {Hill}, {Kirk}, {Ladjelate}, {Marsh}, {Martin},
  {Motte}, {Nguy{\^e}n Luong}, {Pezzuto}, {Roussel}, {Rygl}, {Sadavoy},
  {Schisano}, {Spinoglio}, {Ward-Thompson}, \& {White}}]{konyves2015}
{K{\"o}nyves}, V., {Andr{\'e}}, P., {Men'shchikov}, A., {et~al.} 2015, \aap,
  584, A91

\bibitem[{{Kroupa}(2002)}]{kroupa02}
{Kroupa}, P. 2002, Science, 295, 82

\bibitem[{{Kurtz} {et~al.}(1994){Kurtz}, {Churchwell}, \& {Wood}}]{kurtz00}
{Kurtz}, S., {Churchwell}, E., \& {Wood}, D.~O.~S. 1994, \apjs, 91, 659

\bibitem[{{Lada}(1987)}]{lada87}
{Lada}, C.~J. 1987, in IAU Symposium, Vol. 115, Star Forming Regions, ed.
  M.~{Peimbert} \& J.~{Jugaku}, 1

\bibitem[{{Lada} \& {Adams}(1992)}]{lada92}
{Lada}, C.~J., \& {Adams}, F.~C. 1992, \apj, 393, 278

\bibitem[{{Lada} {et~al.}(1996){Lada}, {Alves}, \& {Lada}}]{lada96}
{Lada}, C.~J., {Alves}, J., \& {Lada}, E.~A. 1996, \aj, 111, 1964

\bibitem[{{Lada} \& {Lada}(2003)}]{lada03}
{Lada}, C.~J., \& {Lada}, E.~A. 2003, \araa, 41, 57

\bibitem[{{Lada} {et~al.}(1993){Lada}, {Young}, \& {Greene}}]{lada93b}
{Lada}, C.~J., {Young}, E.~T., \& {Greene}, T.~P. 1993, \apj, 408, 471

\bibitem[{{Lada} {et~al.}(2006){Lada}, {Muench}, {Luhman}, {Allen}, {Hartmann},
  {Megeath}, {Myers}, {Fazio}, {Wood}, {Muzerolle}, {Rieke}, {Siegler}, \&
  {Young}}]{lada06}
{Lada}, C.~J., {Muench}, A.~A., {Luhman}, K.~L., {et~al.} 2006, \aj, 131, 1574

\bibitem[{{Lada} \& {Lada}(1995)}]{lada95}
{Lada}, E.~A., \& {Lada}, C.~J. 1995, \aj, 109, 1682

\bibitem[{{Liu} {et~al.}(2019){Liu}, {Tan}, {De Buizer}, {Zhang},
  {Beltr{\'a}n}, {Staff}, {Tanaka}, {Whitney}, \& {Rosero}}]{liu19}
{Liu}, M., {Tan}, J.~C., {De Buizer}, J.~M., {et~al.} 2019, \apj, 874, 16

\bibitem[{{L{\'o}pez-Chico} \& {Salas}(2007)}]{lopez07}
{L{\'o}pez-Chico}, T., \& {Salas}, L. 2007, \rmxaa, 43, 155

\bibitem[{{Lucas} {et~al.}(2008){Lucas}, {Hoare}, {Longmore}, {Schr{\"o}der},
  {Davis}, {Adamson}, {Bandyopadhyay}, {de Grijs}, {Smith}, {Gosling},
  {Mitchison}, {G{\'a}sp{\'a}r}, {Coe}, {Tamura}, {Parker}, {Irwin}, {Hambly},
  {Bryant}, {Collins}, {Cross}, {Evans}, {Gonzalez-Solares}, {Hodgkin},
  {Lewis}, {Read}, {Riello}, {Sutorius}, {Lawrence}, {Drew}, {Dye}, \&
  {Thompson}}]{lucas08}
{Lucas}, P.~W., {Hoare}, M.~G., {Longmore}, A., {et~al.} 2008, \mnras, 391, 136

\bibitem[{{Maddox} {et~al.}(2017){Maddox}, {Jarvis}, {Banerji}, {Hewett},
  {Bourne}, {Dunne}, {Dye}, {Eales}, {Furlanetto}, {Maddox}, {Smith}, \&
  {Valiante}}]{maddox17}
{Maddox}, N., {Jarvis}, M.~J., {Banerji}, M., {et~al.} 2017, \mnras, 470, 2314

\bibitem[{{Mateen} {et~al.}(2006){Mateen}, {Hofner}, \& {Araya}}]{mateen06}
{Mateen}, M., {Hofner}, P., \& {Araya}, E. 2006, \apjs, 167, 239

\bibitem[{{McKee} \& {Ostriker}(2007)}]{mckee07}
{McKee}, C.~F., \& {Ostriker}, E.~C. 2007, \araa, 45, 565

\bibitem[{{McKee} \& {Tan}(2003)}]{mckee03}
{McKee}, C.~F., \& {Tan}, J.~C. 2003, \apj, 585, 850

\bibitem[{{Megeath} {et~al.}(1996){Megeath}, {Herter}, {Beichman}, {Gautier},
  {Hester}, {Rayner}, \& {Shupe}}]{megeath96}
{Megeath}, S.~T., {Herter}, T., {Beichman}, C., {et~al.} 1996, \aap, 307, 775

\bibitem[{{Megeath} {et~al.}(2004){Megeath}, {Allen}, {Gutermuth}, {Pipher},
  {Myers}, {Calvet}, {Hartmann}, {Muzerolle}, \& {Fazio}}]{megeath04}
{Megeath}, S.~T., {Allen}, L.~E., {Gutermuth}, R.~A., {et~al.} 2004, \apjs,
  154, 367

\bibitem[{{Meyer} {et~al.}(1997){Meyer}, {Calvet}, \& {Hillenbrand}}]{meyer97}
{Meyer}, M.~R., {Calvet}, N., \& {Hillenbrand}, L.~A. 1997, \aj, 114, 288

\bibitem[{{Miller} \& {Scalo}(1979)}]{miller79}
{Miller}, G.~E., \& {Scalo}, J.~M. 1979, \apjs, 41, 513

\bibitem[{{Molinari} {et~al.}(2016){Molinari}, {Schisano}, {Elia},
  {Pestalozzi}, {Traficante}, {Pezzuto}, {Swinyard}, {Noriega-Crespo}, {Bally},
  {Moore}, {Plume}, {Zavagno}, {di Giorgio}, {Liu}, {Pilbratt}, {Mottram},
  {Russeil}, {Piazzo}, {Veneziani}, {Benedettini}, {Calzoletti}, {Faustini},
  {Natoli}, {Piacentini}, {Merello}, {Palmese}, {Del Grand e}, {Polychroni},
  {Rygl}, {Polenta}, {Barlow}, {Bernard}, {Martin}, {Testi}, {Ali},
  {Andr{\'e}}, {Beltr{\'a}n}, {Billot}, {Carey}, {Cesaroni}, {Compi{\`e}gne},
  {Eden}, {Fukui}, {Garcia-Lario}, {Hoare}, {Huang}, {Joncas}, {Lim}, {Lord},
  {Martinavarro-Armengol}, {Motte}, {Paladini}, {Paradis}, {Peretto},
  {Robitaille}, {Schilke}, {Schneider}, {Schulz}, {Sibthorpe}, {Strafella},
  {Thompson}, {Umana}, {Ward-Thompson}, \& {Wyrowski}}]{molinari16}
{Molinari}, S., {Schisano}, E., {Elia}, D., {et~al.} 2016, \aap, 591, A149

\bibitem[{{Muzerolle} {et~al.}(2004){Muzerolle}, {Megeath}, {Gutermuth},
  {Allen}, {Pipher}, {Hartmann}, {Gordon}, {Padgett}, {Noriega-Crespo},
  {Myers}, {Fazio}, {Rieke}, {Young}, {Morrison}, {Hines}, {Su}, {Engelbracht},
  \& {Misselt}}]{muzerolle04}
{Muzerolle}, J., {Megeath}, S.~T., {Gutermuth}, R.~A., {et~al.} 2004, \apjs,
  154, 379

\bibitem[{{Neugebauer} {et~al.}(1984){Neugebauer}, {Habing}, {van Duinen},
  {Aumann}, {Baud}, {Beichman}, {Beintema}, {Boggess}, {Clegg}, {de Jong},
  {Emerson}, {Gautier}, {Gillett}, {Harris}, {Hauser}, {Houck}, {Jennings},
  {Low}, {Marsden}, {Miley}, {Olnon}, {Pottasch}, {Raimond}, {Rowan-Robinson},
  {Soifer}, {Walker}, {Wesselius}, \& {Young}}]{neugebauer84}
{Neugebauer}, G., {Habing}, H.~J., {van Duinen}, R., {et~al.} 1984, \apjl, 278,
  L1

\bibitem[{{Paron} {et~al.}(2009){Paron}, {Cichowolski}, \&
  {Ortega}}]{Paron2009}
{Paron}, S., {Cichowolski}, S., \& {Ortega}, M.~E. 2009, \aap, 506, 789

\bibitem[{{Persi} \& {Tapia}(2019)}]{persi19}
{Persi}, P., \& {Tapia}, M. 2019, \mnras, 485, 784

\bibitem[{{Peters} {et~al.}(2010){Peters}, {Banerjee}, {Klessen}, {Mac Low},
  {Galv{\'a}n-Madrid}, \& {Keto}}]{peters10}
{Peters}, T., {Banerjee}, R., {Klessen}, R.~S., {et~al.} 2010, \apj, 711, 1017

\bibitem[{{Pezzuto} {et~al.}(2021){Pezzuto}, {Benedettini}, {Di Francesco},
  {Palmeirim}, {Sadavoy}, {Schisano}, {Li Causi}, {Andr{\'e}}, {Arzoumanian},
  {Bernard}, {Bontemps}, {Elia}, {Fiorellino}, {Kirk}, {K{\"o}nyves},
  {Ladjelate}, {Men'shchikov}, {Motte}, {Piccotti}, {Schneider}, {Spinoglio},
  {Ward-Thompson}, \& {Wilson}}]{pezzuto2021}
{Pezzuto}, S., {Benedettini}, M., {Di Francesco}, J., {et~al.} 2021, \aap, 645,
  A55

\bibitem[{{Pilbratt} {et~al.}(2010){Pilbratt}, {Riedinger}, {Passvogel},
  {Crone}, {Doyle}, {Gageur}, {Heras}, {Jewell}, {Metcalfe}, {Ott}, \&
  {Schmidt}}]{pilbratt10}
{Pilbratt}, G.~L., {Riedinger}, J.~R., {Passvogel}, T., {et~al.} 2010, \aap,
  518, L1

\bibitem[{{Poglitsch} {et~al.}(2010){Poglitsch}, {Waelkens}, {Geis},
  {Feuchtgruber}, {Vandenbussche}, {Rodriguez}, {Krause}, {Renotte}, {van
  Hoof}, {Saraceno}, {Cepa}, {Kerschbaum}, {Agn{\`e}se}, {Ali}, {Altieri},
  {Andreani}, {Augueres}, {Balog}, {Barl}, {Bauer}, {Belbachir}, {Benedettini},
  {Billot}, {Boulade}, {Bischof}, {Blommaert}, {Callut}, {Cara}, {Cerulli},
  {Cesarsky}, {Contursi}, {Creten}, {De Meester}, {Doublier}, {Doumayrou},
  {Duband}, {Exter}, {Genzel}, {Gillis}, {Gr{\"o}zinger}, {Henning},
  {Herreros}, {Huygen}, {Inguscio}, {Jakob}, {Jamar}, {Jean}, {de Jong},
  {Katterloher}, {Kiss}, {Klaas}, {Lemke}, {Lutz}, {Madden}, {Marquet},
  {Martignac}, {Mazy}, {Merken}, {Montfort}, {Morbidelli}, {M{\"u}ller},
  {Nielbock}, {Okumura}, {Orfei}, {Ottensamer}, {Pezzuto}, {Popesso},
  {Putzeys}, {Regibo}, {Reveret}, {Royer}, {Sauvage}, {Schreiber}, {Stegmaier},
  {Schmitt}, {Schubert}, {Sturm}, {Thiel}, {Tofani}, {Vavrek}, {Wetzstein},
  {Wieprecht}, \& {Wiezorrek}}]{poglitsch10}
{Poglitsch}, A., {Waelkens}, C., {Geis}, N., {et~al.} 2010, \aap, 518, L2

\bibitem[{{Price} {et~al.}(2001){Price}, {Egan}, {Carey}, {Mizuno}, \&
  {Kuchar}}]{price01}
{Price}, S.~D., {Egan}, M.~P., {Carey}, S.~J., {Mizuno}, D.~R., \& {Kuchar},
  T.~A. 2001, \aj, 121, 2819

\bibitem[{{Qiu} {et~al.}(2008){Qiu}, {Zhang}, {Megeath}, {Gutermuth},
  {Beuther}, {Shepherd}, {Sridharan}, {Testi}, \& {De Pree}}]{qiu08}
{Qiu}, K., {Zhang}, Q., {Megeath}, S.~T., {et~al.} 2008, \apj, 685, 1005

\bibitem[{{Rieke} \& {Lebofsky}(1985)}]{rieke85}
{Rieke}, G.~H., \& {Lebofsky}, M.~J. 1985, \apj, 288, 618

\bibitem[{{Rivera-Ingraham} {et~al.}(2010){Rivera-Ingraham}, {Ade}, {Bock},
  {Chapin}, {Devlin}, {Dicker}, {Griffin}, {Gundersen}, {Halpern}, {Hargrave},
  {Hughes}, {Klein}, {Marsden}, {Martin}, {Mauskopf}, {Netterfield}, {Olmi},
  {Patanchon}, {Rex}, {Scott}, {Semisch}, {Truch}, {Tucker}, {Tucker}, {Viero},
  \& {Wiebe}}]{Rivera2010}
{Rivera-Ingraham}, A., {Ade}, P. A.~R., {Bock}, J.~J., {et~al.} 2010, \apj,
  723, 915

\bibitem[{{Robitaille} {et~al.}(2007){Robitaille}, {Whitney}, {Indebetouw}, \&
  {Wood}}]{robitaille07}
{Robitaille}, T.~P., {Whitney}, B.~A., {Indebetouw}, R., \& {Wood}, K. 2007,
  \apjs, 169, 328

\bibitem[{{Robitaille} {et~al.}(2006){Robitaille}, {Whitney}, {Indebetouw},
  {Wood}, \& {Denzmore}}]{robitaille06}
{Robitaille}, T.~P., {Whitney}, B.~A., {Indebetouw}, R., {Wood}, K., \&
  {Denzmore}, P. 2006, \apjs, 167, 256

\bibitem[{{Robitaille} {et~al.}(2008){Robitaille}, {Meade}, {Babler},
  {Whitney}, {Johnston}, {Indebetouw}, {Cohen}, {Povich}, {Sewilo}, {Benjamin},
  \& {Churchwell}}]{robitaille08}
{Robitaille}, T.~P., {Meade}, M.~R., {Babler}, B.~L., {et~al.} 2008, \aj, 136,
  2413

\bibitem[{{Roy} {et~al.}(2014){Roy}, {Andr{\'e}}, {Palmeirim}, {Attard},
  {K{\"o}nyves}, {Schneider}, {Peretto}, {Men'shchikov}, {Ward-Thompson},
  {Kirk}, {Griffin}, {Marsh}, {Abergel}, {Arzoumanian}, {Benedettini}, {Hill},
  {Motte}, {Nguyen Luong}, {Pezzuto}, {Rivera-Ingraham}, {Roussel}, {Rygl},
  {Spinoglio}, {Stamatellos}, \& {White}}]{roy2014}
{Roy}, A., {Andr{\'e}}, P., {Palmeirim}, P., {et~al.} 2014, \aap, 562, A138

\bibitem[{{Salpeter}(1955)}]{salpeter55}
{Salpeter}, E.~E. 1955, \apj, 121, 161

\bibitem[{{S{\'a}nchez-Portal} {et~al.}(2014){S{\'a}nchez-Portal}, {Marston},
  {Altieri}, {Aussel}, {Feuchtgruber}, {Klaas}, {Linz}, {Lutz}, {Mer{\'\i}n},
  {M{\"u}ller}, {Nielbock}, {Oort}, {Pilbratt}, {Schmidt}, {Stephenson}, \&
  {Tuttlebee}}]{sanchez14}
{S{\'a}nchez-Portal}, M., {Marston}, A., {Altieri}, B., {et~al.} 2014,
  Experimental Astronomy, 37, 453

\bibitem[{{Scalo}(1986)}]{scalo86}
{Scalo}, J.~M. 1986, \fcp, 11, 1

\bibitem[{{Schlegel} {et~al.}(1998){Schlegel}, {Finkbeiner}, \&
  {Davis}}]{schlegel98}
{Schlegel}, D.~J., {Finkbeiner}, D.~P., \& {Davis}, M. 1998, \apj, 500, 525

\bibitem[{{Siess} {et~al.}(2000){Siess}, {Dufour}, \& {Forestini}}]{siess00}
{Siess}, L., {Dufour}, E., \& {Forestini}, M. 2000, \aap, 358, 593

\bibitem[{{Simon} {et~al.}(2001){Simon}, {Jackson}, {Clemens}, {Bania}, \&
  {Heyer}}]{simon01}
{Simon}, R., {Jackson}, J.~M., {Clemens}, D.~P., {Bania}, T.~M., \& {Heyer},
  M.~H. 2001, \apj, 551, 747

\bibitem[{{Solin} {et~al.}(2012){Solin}, {Ukkonen}, \& {Haikala}}]{Solin2012}
{Solin}, O., {Ukkonen}, E., \& {Haikala}, L. 2012, \aap, 542, A3

\bibitem[{{Stern} {et~al.}(2005){Stern}, {Eisenhardt}, {Gorjian}, {Kochanek},
  {Caldwell}, {Eisenstein}, {Brodwin}, {Brown}, {Cool}, {Dey}, {Green},
  {Jannuzi}, {Murray}, {Pahre}, \& {Willner}}]{stern05}
{Stern}, D., {Eisenhardt}, P., {Gorjian}, V., {et~al.} 2005, \apj, 631, 163

\bibitem[{{Vig} {et~al.}(2006){Vig}, {Ghosh}, {Kulkarni}, {Ojha}, \&
  {Verma}}]{vig06}
{Vig}, S., {Ghosh}, S.~K., {Kulkarni}, V.~K., {Ojha}, D.~K., \& {Verma}, R.~P.
  2006, \apj, 637, 400

\bibitem[{{Williams} \& {McKee}(1997)}]{williams97}
{Williams}, J.~P., \& {McKee}, C.~F. 1997, \apj, 476, 166

\bibitem[{{Wood} \& {Churchwell}(1989)}]{wood89}
{Wood}, D. O.~S., \& {Churchwell}, E. 1989, \apjs, 69, 831

\bibitem[{{Wright} {et~al.}(2010){Wright}, {Eisenhardt}, {Mainzer}, {Ressler},
  {Cutri}, {Jarrett}, {Kirkpatrick}, {Padgett}, {McMillan}, {Skrutskie},
  {Stanford}, {Cohen}, {Walker}, {Mather}, {Leisawitz}, {Gautier}, {McLean},
  {Benford}, {Lonsdale}, {Blain}, {Mendez}, {Irace}, {Duval}, {Liu}, {Royer},
  {Heinrichsen}, {Howard}, {Shannon}, {Kendall}, {Walsh}, {Larsen}, {Cardon},
  {Schick}, {Schwalm}, {Abid}, {Fabinsky}, {Naes}, \& {Tsai}}]{wright10}
{Wright}, E.~L., {Eisenhardt}, P.~R.~M., {Mainzer}, A.~K., {et~al.} 2010, \aj,
  140, 1868

\bibitem[{{Wu} {et~al.}(2019){Wu}, {Reid}, {Sakai}, {Dame}, {Menten},
  {Brunthaler}, {Xu}, {Li}, {Ho}, {Zhang}, {Rygl}, \& {Zheng}}]{wu19}
{Wu}, Y.~W., {Reid}, M.~J., {Sakai}, N., {et~al.} 2019, \apj, 874, 94

\bibitem[{{Zinnecker} {et~al.}(1993){Zinnecker}, {McCaughrean}, \&
  {Wilking}}]{zinnecker93}
{Zinnecker}, H., {McCaughrean}, M.~J., \& {Wilking}, B.~A. 1993, in Protostars
  and Planets III, ed. E.~H. {Levy} \& J.~I. {Lunine}, 429--495

\bibitem[{{Zinnecker} \& {Yorke}(2007)}]{zinnecker07}
{Zinnecker}, H., \& {Yorke}, H.~W. 2007, \araa, 45, 481

\end{thebibliography}
\newgeometry{margin=1.2cm} 
\renewcommand{\tabcolsep}{0.08cm}
\begin{landscape}
{\fontsize{7.7}{7}\selectfont 
\begin{center}
\begin{longtable}{l l l *{13}{c} l}
\caption[]{NIR and MIR photometric data of selected YSOs} \label{tab:NIR} \\

\hline\hline\noalign{\smallskip}

ID	&	 $\alpha$(2000)			&	$\delta$(2000)			&	J			&	H			&	K			&	[3.6]\,$\mu$ m			&	[4.5]\,$\mu$ m			&	[5.8]\,$\mu$ m			&	[8.0]\,$\mu$ m			&	[24]\,$\mu$ m			&	W1			&	W2			&	W3			&	W4			&	$\alpha$	&	Class$_{Final}$	\\
	&	(hh mm ss)  			&	(dd mm ss) 			&	(mag)			&	(mag)			&	(mag)			&	(mag)			&	(mag)			&	(mag)			&	(mag)			&	(mag)			&	(mag)			&	(mag)			&	(mag)			&	(mag)			&		&		\\
\hline\noalign{\smallskip}																							
(1)     &               (2)             &               (3)                 &            (4)             &               (5)             &               (6)             &               (7)             &               (8)             &               (9)             &               (10)            &               (11)            &               (12)            &       (13) & (14) & (15) & (16) & (17)   \\				

\endfirsthead

\multicolumn{17}{c}%
{{\bfseries \tablename\ \thetable{} -- continued from previous page}} \\
\hline\hline\noalign{\smallskip}
ID	&	 $\alpha$(2000)			&	$\delta$(2000)			&	J			&	H			&	K			&	[3.6]\,$\mu$ m			&	[4.5]\,$\mu$ m			&	[5.8]\,$\mu$ m			&	[8.0]\,$\mu$ m			&	[24]\,$\mu$ m			&	W1			&	W2			&	W3			&	W4			&	$\alpha$	&	Class$_{Final}$	\\
	&	(hh mm ss)  			&	(dd mm ss) 			&	(mag)			&	(mag)			&	(mag)			&	(mag)			&	(mag)			&	(mag)			&	(mag)			&	(mag)			&	(mag)			&	(mag)			&	(mag)			&	(mag)			&		&		\\
\hline\noalign{\smallskip}																							
(1)     &               (2)             &               (3)                 &            (4)             &               (5)             &               (6)             &               (7)             &               (8)             &               (9)             &               (10)            &               (11)            &               (12)            &       (13) & (14) & (15) & (16) & (17)   \\				

\hline\noalign{\smallskip}
\endhead
\hline 
\endfoot
\hline \hline
\endlastfoot

\hline \noalign{\smallskip}												\multicolumn{17}{c}{IRAS\,19111+1048}\\								\hline \noalign{\smallskip}							
IR* 	&	19	13	27.34	&	10	53	32.45	&	16.8	$\pm$ 	0.014	&	15.3	$\pm$ 	0.008	&	14.0	$\pm$ 	0.005	&		$-$		&		$-$		&		$-$		&		$-$		&		$-$		&		$-$		&		$-$		&		$-$		&		$-$		&	$-$	&	ClassII	\\
1	&	19	13	22.64	&	10	54	16.43	&	18.0	$\pm$ 	0.041	&	15.5	$\pm$ 	0.010	&	14.4	$\pm$ 	0.007	&	13.4	$\pm$ 	0.073	&	13.2	$\pm$ 	0.171	&		$-$		&		$-$		&		$-$		&		$-$		&		$-$		&		$-$		&		$-$		&	$-$	&	ClassII	\\
2	&	19	13	23.28	&	10	54	34.45	&	17.9	$\pm$ 	0.040	&	15.3	$\pm$ 	0.008	&	14.0	$\pm$ 	0.005	&	13.0	$\pm$ 	0.082	&	12.7	$\pm$ 	0.114	&		$-$		&		$-$		&		$-$		&		$-$		&		$-$		&		$-$		&		$-$		&	$-$	&	ClassII	\\
3	&	19	13	23.6	&	10	54	17.61	&	19.3	$\pm$ 	0.138	&	15.9	$\pm$ 	0.014	&	14.4	$\pm$ 	0.007	&	13.3	$\pm$ 	0.081	&	13.1	$\pm$ 	0.148	&		$-$		&		$-$		&		$-$		&		$-$		&		$-$		&		$-$		&		$-$		&	$-$	&	ClassII	\\
4	&	19	13	23.89	&	10	54	9.38	&	17.9	$\pm$ 	0.037	&	17.0	$\pm$ 	0.037	&	16.5	$\pm$ 	0.048	&		$-$		&		$-$		&		$-$		&		$-$		&		$-$		&		$-$		&		$-$		&		$-$		&		$-$		&	$-$	&	ClassII	\\
5	&	19	13	23.94	&	10	54	0.87	&	18.2	$\pm$ 	0.052	&	15.7	$\pm$ 	0.012	&	14.5	$\pm$ 	0.008	&	13.6	$\pm$ 	0.117	&	13.5	$\pm$ 	0.140	&		$-$		&		$-$		&		$-$		&		$-$		&		$-$		&		$-$		&		$-$		&	$-$	&	ClassI	\\
6	&	19	13	24.36	&	10	53	53.01	&	19.0	$\pm$ 	0.106	&	16.2	$\pm$ 	0.018	&	14.9	$\pm$ 	0.011	&		$-$		&		$-$		&		$-$		&		$-$		&		$-$		&		$-$		&		$-$		&		$-$		&		$-$		&	$-$	&	ClassI	\\
\hline
\end{longtable}
\end{center}
}
(1) - ID number in final list of YSOs, (2),(3) - position taken from the UKIDSS survey, (4)–(15) - apparent magnitudes with errors, (16) - $\alpha_{IRAC}$, (17) - classification of YSOs according to c-c diagrams and $\alpha_{IRAC}$. W1–W4 are four bands of WISE survey. *- IRAS\,19111+1048.
\end{landscape}
\restoregeometry

\newgeometry{margin=1.2cm} 
\renewcommand{\tabcolsep}{0.15cm}
{\fontsize{8}{7}\selectfont 
\begin{center}
\begin{longtable}{ *{9}{l} }
\caption[]{FIR photometric data of selected YSOs} \label{tab:FIR} \\
\hline\hline\noalign{\smallskip}
ID	&	 $\alpha$(2000)			&	$\delta$(2000)			&	[70]\,$\mu$m			&	[160]\,$\mu$m			&	[250]\,$\mu$m			&	[350]\,$\mu$m			&	[500]\,$\mu$m			&		Class$_{Final}$	\\
	&	(hh mm ss)  			&	(dd mm ss) 			&	(mJy)			&	(mJy)			&	(mJy)			&	(mJy)			&	(mJy)			&			\\
\hline\noalign{\smallskip}
(1)     &               (2)             &               (3)                 &            (4)             &               (5)             &               (6)             &               (7)             &               (8)             &               (9) \\

\hline\noalign{\smallskip}
\multicolumn{9}{c}{IRAS\,19111+1048}\\
\hline \noalign{\smallskip}
IR* 	&	19	13	27.34	&	10	53	32.45	&		$-$		&	1326590	$\pm$ 	1069	&		$-$		&		$-$		&		$-$		&	ClassII	\\
20	&	19	13	25.30	&	10	54	8.29	&	14369	$\pm$ 	335	&		$-$		&		$-$		&		$-$		&		$-$		&		ClassII	\\
35	&	19	13	26.22	&	10	54	39.42	&	122665	$\pm$ 	589	&		$-$		&		$-$		&		$-$		&		$-$		&	ClassII	\\
\hline
\end{longtable}
\end{center}
}
(1) - ID number in final list of YSOs, (2),(3) - position taken from the UKIDSS survey and HiGAL 160\,$\mu$m 
(4)–(8) - measured fluxes with errors,  (9) - classification of YSOs. *- IRAS\,19111+1048.

\restoregeometry

\newgeometry{margin=1.2cm} 
\renewcommand{\tabcolsep}{0.15cm}
{\fontsize{8}{7}\selectfont 
\begin{center}
\begin{longtable}{l l l *{4}{c} l *{4}{c} l}
\caption[]{Parameters derived from \citet{robitaille07} models SED fitting.} \label{tab:SED} \\
\hline \hline \noalign{\smallskip}

ID	&	N$_{d}$ 	&	N$_{fit}$	&	A$_v$			&	Distance			&	Stellar age			&	Stellar mass			&	Temperature			&	M$_{disk}$			&	MdotE			&	MdotD			&	L$_{Total}$			&			Class	\\
	&		&		&	(mag)			&	(kpc)			&	(Log)(yr)			&	(M$_{\odot}$)		&	(K)			&	 (Log)(M$_{\odot} $)			&	 (Log)(M$_{\odot} yr^{-1}$)			& (Log)(M$_{\odot} yr^{-1}$)			&	(Log)(L$_{\odot}$)					&	\\
\hline \noalign{\smallskip}														
\hline\noalign{\smallskip}
(1)     &               (2)             &               (3)                 &            (4)             &               (5)             &               (6)             &               (7)             &               (8)             &               (9)             &               (10)            &               (11)            &               (12)            &       (13)  \\

\endfirsthead

\multicolumn{12}{c}%
{{\bfseries \tablename\ \thetable{} -- continued from previous page}} \\
\hline\hline\noalign{\smallskip}
ID	&	N$_{d}$ 	&	N$_{fit}$	&	A$_v$			&	Distance			&	Stellar age			&	Stellar mass			&	Temperature			&	M$_{disk}$			&	MdotE			&	MdotD			&	L$_{Total}$			&			Class	\\
	&		&		&	(mag)			&	(kpc)			&	(Log)(yr)			&	(M$_{\odot}$)		&	(K)			&	 (Log)(M$_{\odot} $)			&	 (Log)(M$_{\odot} yr^{-1}$)			& (Log)(M$_{\odot} yr^{-1}$)			&	(Log)(L$_{\odot}$)					&	\\
\hline \noalign{\smallskip}														
\hline\noalign{\smallskip}
(1)     &               (2)             &               (3)                 &            (4)             &               (5)             &               (6)             &               (7)             &               (8)             &               (9)             &               (10)            &               (11)            &               (12)            &       (13)  \\				

\hline\noalign{\smallskip}
\endhead
\hline 
\endfoot
\hline \hline
\endlastfoot

\hline \noalign{\smallskip}
\multicolumn{13}{c}{IRAS\,19111+1048}\\								\hline \noalign{\smallskip}						
IR*	&	3	&	4	&	12.0	$\pm$	4.6	&	8.7	$\pm$	2.2	&	6.40	$\pm$	6.07	&	9.4	$\pm$	4.3	&	22928	$\pm$	10771	&	-2.9	$\pm$	-2.9	&	-8.4	$\pm$	-8.4	&	-7.8	$\pm$	-7.7	&	3.8	$\pm$	3.6	&	II	\\																																																														
1	&	5	&	260	&	15.0	$\pm$	2.1	&	7.4	$\pm$	1.3	&	5.15	$\pm$	4.38	&	3.2	$\pm$	0.4	&	4542	$\pm$	622	&	-2.3	$\pm$	-2.6	&	-4.1	$\pm$	-4.5	&	-6.7	$\pm$	-7.0	&	1.8	$\pm$	0.9	&	II	\\																																																														
2	&	5	&	182	&	16.8	$\pm$	3.2	&	7.8	$\pm$	1.4	&	5.10	$\pm$	4.51	&	4.7	$\pm$	1.0	&	4677	$\pm$	861	&	-2.4	$\pm$	-2.5	&	-4.1	$\pm$	-4.5	&	-7.0	$\pm$	-7.3	&	2.1	$\pm$	1.4	&	II	\\																																																														
3	&	5	&	131	&	21.1	$\pm$	3.4	&	7.5	$\pm$	1.3	&	5.04	$\pm$	4.53	&	4.7	$\pm$	0.7	&	4619	$\pm$	568	&	-1.7	$\pm$	-1.6	&	-4.0	$\pm$	-3.9	&	-7.0	$\pm$	-6.8	&	2.1	$\pm$	1.5	&	II	\\																																																														
5	&	5	&	246	&	15.4	$\pm$	2.7	&	7.5	$\pm$	1.4	&	5.16	$\pm$	4.41	&	3.1	$\pm$	0.5	&	4538	$\pm$	788	&	-2.4	$\pm$	-2.5	&	-4.1	$\pm$	-4.4	&	-6.9	$\pm$	-7.1	&	1.8	$\pm$	1.0	&	I	\\																																																														
7	&	4	&	7	&	10.6	$\pm$	1.5	&	6.9	$\pm$	1.3	&	6.69	$\pm$	6.54	&	4.5	$\pm$	0.9	&	10808	$\pm$	4290	&	-3.3	$\pm$	-3.0	&	-5.1	$\pm$	-4.9	&	-8.7	$\pm$	-8.4	&	2.4	$\pm$	2.3	&	I	\\																																																														
10	&	5	&	198	&	16.2	$\pm$	1.9	&	7.7	$\pm$	1.3	&	4.99	$\pm$	3.92	&	3.5	$\pm$	0.3	&	4497	$\pm$	491	&	-2.7	$\pm$	-3.1	&	-3.9	$\pm$	-4.6	&	-6.8	$\pm$	-7.4	&	1.9	$\pm$	0.9	&	II	\\
\hline
\end{longtable}
\end{center}
}
(1) - ID number in final list of YSOs, (2) - number of fitting models, (3) - number of input fluxes, (4)-(12) - the weighted means and the standard deviations of parameters obtained by SED fitting tool for all models with best $\chi^{2}-\chi^{2}_{best}$\,<\,3N, 
(13) - classification of YSOs. *-IRAS\,19111+1048.
\restoregeometry

\end{document}